\documentclass[12pt]{iopart}
\pdfoutput=1

\usepackage{iopams}
\usepackage{bbm}
\usepackage{bm}
\usepackage[pdftex]{graphicx}
\usepackage{hyperref}

\expandafter\let\csname equation*\endcsname\relax
\expandafter\let\csname endequation*\endcsname\relax

\usepackage{amsmath}

 \usepackage[caption=false]{subfig}
 \usepackage{floatrow}
 \usepackage{longtable}
 \usepackage{multirow}
\usepackage{adjustbox}
\DeclareMathOperator{\pr}{\text{pr}}
\DeclareMathOperator{\dd}{\text{d}\!}
\DeclareMathOperator{\Log}{\text{Log}}

\usepackage{dcolumn}

\newcommand{\strike}[1]{\ifmmode\setbox0\hbox{$#1$}%
\else
\setbox0\hbox{#1}%
\fi
\makebox[\the\wd0][c]{%
\rule[0.48\ht0]{0.5\wd0}{0.25pt}}\hspace*{-\the\wd0}#1}

\usepackage{color}

\begin{document}

\title[]{Rare desynchronization events in power grids: On data implementation and dimensional reductions}

\author{Tim Ritmeester and Hildegard Meyer-Ortmanns}

\address{School of Science, Jacobs University Bremen, Campus Ring 1, 28759 Bremen, Germany}
\ead{t.ritmeester@jacobs-university.de}
\ead{h.ortmanns@jacobs-university.de}
\vspace{10pt}
\begin{indented}
\item[]November 2022
\end{indented}

\begin{abstract}
We discuss the frequency of desynchronization events in power grids for realistic data input. We focus on the role of time correlations in the fluctuating power production and propose a new method for implementing colored noise that reproduces non-Gaussian data by means of cumulants of data increment distributions. Our desynchronization events are caused by overloads. We extend known and propose different methods of dimensional reduction to considerably reduce the high-dimensional phase space and to predict the rare desynchronization events with reasonable computational costs. The first method splits the system into two areas, connected by heavily loaded lines, and treats each area as a single node. The second method considers a separation of the timescales of power fluctuations and phase angle dynamics and completely disregards the latter. The fact that this separation turns out to be justified, albeit only to exponential accuracy in the strength of fluctuations, means that the number of rare events does not sensitively depend on inertia or damping for realistic heterogeneous parameters and long correlation times. Neither does the number of desynchronization events automatically increase with non-Gaussian fluctuations in the power production as one might have expected. On the other hand, the analytical expressions for the average time to desynchronization depend sensitively on the finite correlation time of the fluctuating power input.
\end{abstract}

\section{Introduction}

Power grids are confronted with various sources of stochastic fluctuations. Fluctuations refer to production, consumption, and prices on the energy market, where supply and demand are traded to cure imbalances between forecast and actual needs of energy supply. Variations in supply and demand have always existed, but in view of decarbonization, the increase in the contribution of renewable energies has dramatically increased fluctuations due to larger and more localized variability of wind and solar energy \cite{kabouris_impacts_2009}. As a result, power grids become increasingly affected by collective, nonlinear phenomena, which are less predictable in space and time, more heterogeneous and less well controllable. For a recent review see \cite{hilde}.

Wind and solar energies are currently fast-growing energy sectors in Europe. In addition, more renewables in the grid go along with stronger fluctuations, less inertia and less damping. How fluctuations induced by renewables challenge the stability and quality of electrical power grids was analyzed in \cite{katrin}, addressing the regime in which desynchronization events are not rare.
Here, feed-in fluctuations were modelled with realistic features of non-vanishing temporal correlations, the Kolmogorov power spectrum and intermittent increments. Less inertia and damping are expected to further challenge the grid stability. The impact of increased fluctuations, reduced inertia and reduced damping on the grid stability was studied in \cite{schafer_escape_2017}. The fluctuations were implemented as Gaussian white noise and assumed to be small. The analysis was based on employing Kramer's escape rate. Accordingly, inertia has a minor effect on the escape rate, entering the prefactor of the exponential, while the damping occurs in the exponent with a stronger effect. The escape routes were shown to typically proceed in the vicinity of saddle fixed points with a low potential barrier, caused by a single overloaded link. \\
As shown in \cite{manik_supply_2014}, a stable fixed point of the swing equations can be only lost via an inverse saddle node bifurcation. As long as the phase differences along all transmission lines are smaller or equal $\pi/2$, which is the case under normal grid operation \cite{machowski_power_2008,ENTSOEreport}, the loss of a stable fixed point is equivalent to an overload of one or more transmission lines. It is this case that we consider later in this paper. However, if the condition on the phase differences is violated, the bifurcation of a fixed point is generally not associated with an overload.\\
Histograms of power increments of wind and solar data show clearly non-Gaussian tails. Non-Gaussian fluctuations were considered in \cite{hindes_network_2019} to determine rare events of power grid desynchronization, where the model was based on the swing equations under colored noise. The computed rate of desynchronization showed that higher moments of noise enter at specific powers of the coupling. Their effect is either increasing or decreasing the rate of desynchronization events, depending on how the noise statistics match the statistics of the Fiedler mode, which is the slowest mode in the system. Their work makes use of a WKB-approach to determine the average time to desynchronization. For earlier work that uses the WKB approach for classical stochastic systems see, for example, \cite{dykman1,dykman2}.
We will also employ the WKB-approach, but use different versions of dimensional reduction for obtaining our predictions.\\
When strong fluctuations in the production propagate through the grid, non-Gaussian tails in the distribution of local grid-frequency data have been observed, which themselves may induce large-scale blackouts in the worst case. Here the authors of \cite{tyloo_finite-time_2022} focused on the role of time correlations in the fluctuations. According to their results, noise disturbances with finite skewness and positive kurtosis propagate through the entire grid when the noise correlation time is larger than the network's intrinsic time scales. Otherwise, for rapidly fluctuating noise, the disturbances decay over short distances. We will also focus on the role of time correlations in the non-Gaussian noise, but pursue their impact on the average time to desynchronization events. \\
The paper is organized as follows. In Sec.~\ref{sec2} we present the model in terms of the swing equations with stochastic power input, described as non-Gaussian colored noise to reproduce in particular real data of wind power increment distributions. As network we choose a coarse-grained version of the Brazilian grid with prototypical values of inertia and damping, while the power production is artificially increased to simulate a heavily loaded grid at risk of desynchronization. Examples of typical histograms of power increments are shown, as measured from wind and solar data. Sec.~\ref{sec3} is devoted to the data implementation. We propose a new approach, the so-called Fourier-method, and compare it with a compound Poisson process, previously used, for example, in \cite{hindes_network_2019}. The Fourier-method is analytically more convenient, and more accurately reproduces renewable generation data. Both approaches can lead to different predictions of the desynchronization times, depending on how  the data are implemented on the time scale of one integration step. In Sec.~\ref{sec4} we derive various versions of dimensional reduction of the equations of motion, first applied to the swing equations, next to Hamilton's equations of motion that show up in the WKB-approach. We reduce the phase space of Hamilton's equations for the full grid to twelve dimensions (in the synchronized subgraph approximation), and to two dimensions (in the overload approximation). In Sec.~\ref{sec5} we explain the methods of finding the solutions. Sec.~\ref{sec: results} contains a comparison of the results for the average desynchronization times for different versions of data implementation and approximation schemes. We demonstrate the accuracy of these approximations by comparing their results to those obtained for the full system. We also discuss the role of the skewness of the power increment distributions. Though derived differently, our results confirm that the number of rare desynchronization events may increase or decrease due to more renewables, depending on where the fluctuations enter the grid. Moreover, we find that neither less inertia nor less damping endangers the grid stability per se with respect to desynchronization. The role of time correlations in the fluctuations is clearly non-negligible, correlation times are related to the time scales of inertia and damping.  Sec.~\ref{sec7} contains our summary and conclusions.
\section{The model}\label{sec2}
After introducing the model (Sec.~\ref{subsec21}) and discussing the role of time correlations in Sec.~\ref{sec: time_correlations}, we give some details about the Brazilian grid as a grid that shares prototypical features in view of realistic data for production, consumption and inertia (Sec.~\ref{subsec22}), followed by examples of real data sets (Sec.~\ref{subsec23}).

\subsection{Swing equation under realistic data input}\label{subsec21}
If one neglects resistive power losses, assumes constant voltage magnitudes and phase velocities much smaller than the grid frequency ($\vert\dot{\phi}\vert\ll\omega$), the dynamics of the voltage phase angles $\bm{\phi}$ are modeled by the swing equations \cite{manik_supply_2014,machowski_power_2008,nishikawa_comparative_2015}:
\begin{alignat}{1}
    \frac{\dd \phi_i}{\dd t} &=v_i \nonumber \\
    M_i \frac{\dd v_i}{\text{d}t} + \gamma_i v_i & =  P_i + p_i(t) + \sum_j K_{ij} \sin(\phi_j - \phi_i), \label{eq: swing_dynamics}
\end{alignat}
Here $M_i$ and $\gamma_i$ are respectively the inertia and the damping constants of the oscillators, $K_{ij}$ is the susceptance of the power line between nodes $i$ and $j$ (equal to $0$ if nodes $i$ and $j$ are not adjacent). $K_{ij} \sin(\phi_j - \phi_i)$ equals the flow $f_{ij}$ of power from node $j$ to node $i$. For a detailed derivation of why the swing equations are representative for modeling the grid dynamics (without including the voltage dynamics) we refer to the work of \cite{manik_supply_2014,machowski_power_2008,nishikawa_comparative_2015}. The constants $P_i$ are deterministic power injections with $\sum_i P_i = 0$, and $p_i(t)$ are unforeseen fluctuations in power input. Fluctuations in power input are both non-Gaussian and colored \cite{anvari_short_2016,lopes_use_2012,haehne_propagation_2019}.

A convenient noise model which incorporates both of these properties is the Langevin-like dynamics (following \cite{hindes_network_2019, kampen_stochastic_1992}):
\begin{alignat}{1}
    \frac{\text{d}p_i(t)}{\text{d}t} & = - \alpha p_i(t) + \sigma_i \cdot \xi_i(t), \label{eq: p_dynamics}
\end{alignat}
where $\xi_i(t)$ is stationary (generally non-Gaussian) white noise with $\langle \xi_i(t) \xi_j(t') \rangle = \delta_{ij} \delta(t - t')$, and where $\sigma_i$ controls the strength of the noise. As  the solution to Eq.~\ref{eq: p_dynamics} is $p_i(t) = \sigma_i \int_{-\infty}^t \dd t' \, \exp(-\alpha (t - t')) \xi_i(t')$, Eq.~\ref{eq: p_dynamics} leads to exponentially decaying time correlations $\frac{\langle p_i(t)p_i(t') \rangle}{\langle p_i(t)p_i(t) \rangle} = \exp\big(- \alpha(t - t')\big)$, so $1/\alpha$ sets the correlation time of the fluctuations. This exponential decay may not really reflect realistic time correlations in feed-in power fluctuations, but provides a tractable approach to pursue the impact of a finite (long) correlation time in contrast to a vanishing one as for white noise. Furthermore this model for the noise allows also to reproduce the measured increment distributions of power input as generated by real wind fluctuations, though their power spectrum is not reproduced. The authors of \cite{katrin} found that in the regime where desynchronization events are common, non-Gaussianity has only a small effect on desynchronization times; our approach will clarify the more important role of non-Gaussian effects in the regime where such events are rare.

\subsection{Role of time correlations in  power fluctuations} \label{sec: time_correlations}
As our focus is on the impact of realistic parameter choices on the grid dynamics, the role of the correlation time of fluctuations (here of the power production, with time correlations parameterized by $1/\alpha)$ should be considered. In the per-unit system power is dimensionless, and $M_i/P_i$ and $\gamma_i/P_i$ have units of $\text{time}^2$ and $\text{time}$, respectively. The influence of the correlation time $1/\alpha$ on the dynamics of the swing equations can be made explicit if we consider a rescaled time $\tilde{t} \equiv \alpha t$, and define:
\begin{alignat}{1}
    \tilde{\phi}_i(\tilde{t}) &\equiv \phi_i(\tilde{t}/\alpha) = \phi_i(t) \,,\\
    \tilde{p}_i(\tilde{t}) &\equiv p_i(\tilde{t}/\alpha) = p_i(t) \,,\\
    \tilde{\xi}_i(\tilde{t}) &\equiv \alpha^{-1/2} \xi_i(\tilde{t}/\alpha) = \alpha^{-1/2}\xi_i(t) \,,
\end{alignat}
then using $\frac{\text{d} \tilde{\phi}_i(\tilde{t})}{\text{d}\tilde{t}} 
= \frac{1}{\alpha} \frac{\text{d}\phi_i(t)}{\text{d}t}$ and $\frac{\text{d}^2 \tilde{\phi}_i(\tilde{t})}{\text{d}\tilde{t}^2} 
= \frac{1}{\alpha^2} \frac{\text{d}^2\phi_i(t)}{\text{d}t^2}$, the swing equations (Eq.~\ref{eq: swing_dynamics}) become:
\begin{alignat}{1}
    (\alpha^2 {M}_i) \cdot \frac{\text{d}^2 \tilde{\phi}_i}{\text{d}\tilde{t}^2} + (\alpha {\gamma}_i) \cdot \frac{\text{d} \tilde{\phi}_i}{\text{d}\tilde{t}} & =  P_i + \tilde{p}_i(\tilde{t}) + \sum_j K_{ij} \sin(\tilde{\phi}_j - \tilde{\phi}_i) \,.\label{eq:10}
\end{alignat}

Furthermore, using that $\frac{\text{d}\tilde{p}_i(\tilde{t})}{\text{d}\tilde{t}} 
 = \frac{1}{\alpha}\frac{\text{d}{p}_i(t)}{\text{d}t}$ and $\delta(\tilde{t}_1 - \tilde{t}_2) = \delta(\alpha ({t}_1 - {t}_2)) = \frac{1}{\alpha }\delta({t}_1 - {t}_2)$ turns Eq.~\ref{eq: p_dynamics} into:
\begin{alignat}{1}
    \frac{\text{d}\tilde{p}_i}{\text{d} \tilde{t}} = - \tilde{p}_i + (\alpha^{-1/2} \sigma_i) \cdot \tilde{\xi}_i(\tilde{t})
\end{alignat}
with $\langle \langle \tilde{\xi}_i(\tilde{t})\tilde{\xi}_i(\tilde{t}') \rangle\rangle = \delta(\tilde{t} - \tilde{t'})$ and  $\langle \langle \tilde{\xi}_i(\tilde{t}_1) \dots \tilde{\xi}_i(\tilde{t}_n) \rangle \rangle = \big(\alpha^{\frac{n}{2} -1}{\Gamma}_n^{\xi_i} \big) \cdot \delta(\tilde{t}_1 - \tilde{t}_2) \dots   \delta(\tilde{t}_1 - \tilde{t}_n)$, where $\langle\langle...\rangle\rangle$ denote the cumulants (Sec.~\ref{sec3}, \cite{kampen_stochastic_1992, gardiner_stochastic_2009}), here of $\tilde{\xi}_i$.
Therefore, setting the correlation time to $1$ by rescaling time according to $t \rightarrow \alpha t$ sends $M_i \rightarrow \alpha^2 M_i$, $\gamma_i \rightarrow \alpha \gamma_i$, $\sigma_i^2 \rightarrow \sigma_i^2/\alpha$ and ${\Gamma}_n^{\xi_i} \rightarrow \alpha^{\frac{n}{2} -1}{\Gamma}_n^{\xi_i}$, with the combinations $\alpha^2 M_i$ and $\alpha \gamma_i$ invariant. Increasing the timescale $1/\alpha$ of the noise is thus equivalent to decreasing the inertia $\bm{M}$ and damping $\bm{\gamma}$. The right-hand side of Eq.~\ref{eq: p_dynamics} is, in the per unit system, dimensionless, and therefore invariant under a change in $1/\alpha$. The rescaled strength of noise increases for increasing $1/\alpha$, while the higher order cumulants of non-Gaussian noise get increasingly suppressed. This means, for non-Gaussian noise, the noise terms become closer to Gaussian if $1/\alpha > 1\,\text{s}$, and vice-versa. In Sec.~\ref{sec: results} this will be manifest in the results for the average time to desynchronization. \\
A further discussion comparing $M_i$ and $\gamma_i$ to the timescale of the noise is given in \cite{tyloo_finite-time_2022}, where it is argued that typical timescales of fluctuations are so long ($1/\alpha = 100 - 300 \;\text{s}$) that the impact of $\bm{M}$ and $\bm{\gamma}$ on the propagation of non-Gaussian disturbances can typically be neglected. The results of \cite{tyloo_finite-time_2022} are for typical fluctuations; our results are derived for rare large fluctuations, thus complementing their results.
\subsection{The Brazilian grid as a prototypical case}\label{subsec22}
\begin{figure}
    \centering
    \subfloat[]{\includegraphics[width = 0.3 \textwidth]{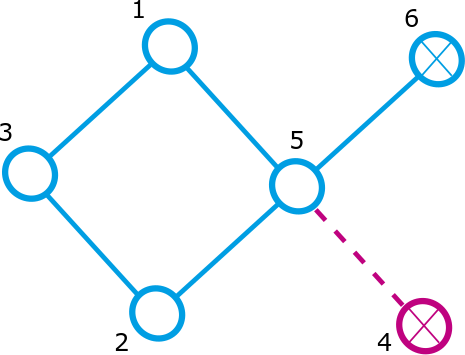}}
    \subfloat[]{\hspace{0.5cm}\includegraphics[width = 0.3 \textwidth]{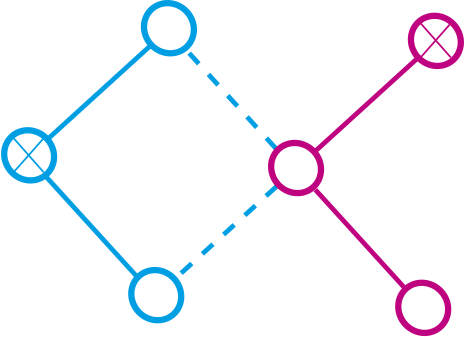}}
    \subfloat[]{\hspace{0.5cm} \includegraphics[width = 0.3 \textwidth]{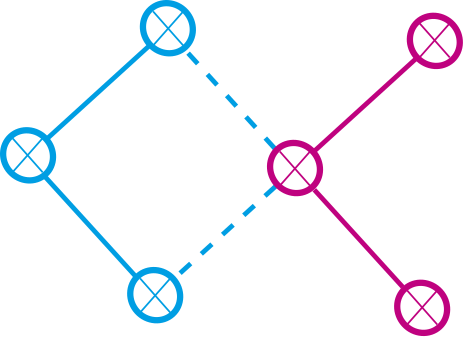}}
    \caption{a) Coarse-grained network  based on the Brazilian power grid \cite{birchfield_grid_2017, xu_creation_2017, xu_modeling_2018} (Sec.~\ref{subsec22}). The dotted lines are the most heavily loaded lines, one in (a), two in (b) and (c); after a desynchronization event the system splits into two areas (with blue and pink nodes, respectively). The two crossed nodes in (a) and (b) contain large power generation/load, to which we add stochastic fluctuations. In c) stochastic fluctuations (with the same variance) are added to all nodes. The networks in (b) and (c) differ from a) as the values of $P_i$, $\gamma_i$ and $M_i$ for the $3$rd and $4$th nodes are swapped. }
    \label{fig: brazilian_network}
\end{figure}

\begin{figure}
    \centering
    \subfloat[]{\includegraphics[width = 0.5 \textwidth]{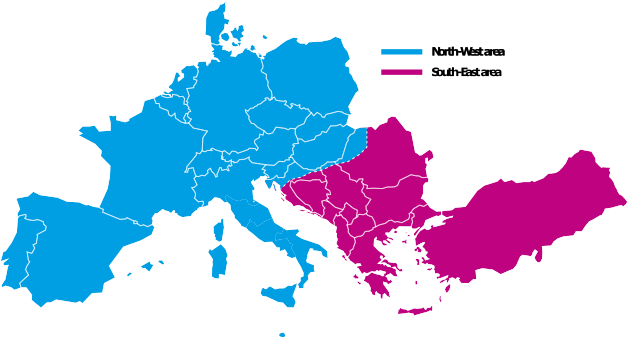}}
    \subfloat[]{\includegraphics[width = 0.5 \textwidth]{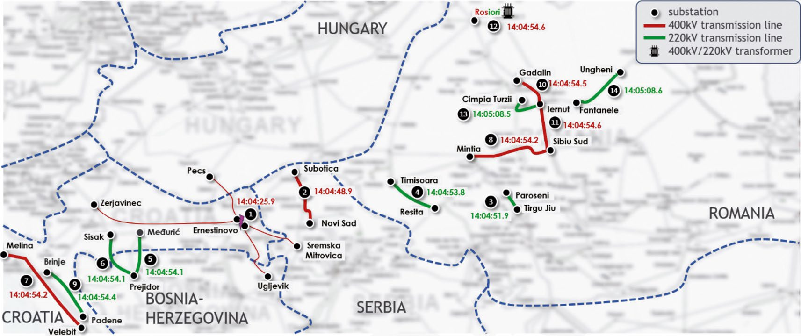}}
    \caption{A system split that occurred in Europe in January 2021 \cite{ENTSOEreport}. The two areas, indicated in (a), lost synchronization between each other due to an overload of the red and green lines connecting the two areas indicated in (b), with the phase angles on either end of the lines reaching a difference of $\frac{\pi}{2}$. Figures included (with permission) from \cite{ENTSOEreport}.}
    \label{fig: system_split}
\end{figure}

As a realistic test case for the stability of the fixed point under stochastic fluctuation we use the coarse-grained data on the Brazilian power network presented in \cite{birchfield_grid_2017, xu_creation_2017, xu_modeling_2018}. The network has $6$ nodes, realistic and spatially heterogeneous parameters, and contains both synchronous generators/loads and inductive generators/loads. The network topology is shown in Fig.~\ref{fig: brazilian_network}. The networks of (b) and (c) have the values of $P_i$, $\gamma_i$ and $M_i$ of nodes 3 and 4 swapped as compared to (a). The networks contain two nodes with high power generation/consumption; nodes $4$ and $6$ in (a) and nodes $3$ and $6$ in (b) and (c). In (a) and (b) we artificially add stochastic fluctuations $\bm{\xi}$ to the power input at these two nodes (crossed in Fig.~\ref{fig: brazilian_network}) to simulate a fraction of the input being fluctuating rather than constant.  In (c) we add power fluctuations to all nodes. The dotted lines are the most heavily loaded lines (with the smallest $\frac{K_{ij} - f_{ij}}{K_{ij}}$), one in (a), two in (b) and (c). If the load on the grid is increased, the system undergoes a bifurcation in which the system splits into two areas (with blue and pink nodes, respectively) on either end of the line(s), which remain synchronized internally (Sec.~\ref{subsec41}). It should be noticed that a splitting into two areas (the blue and pink ones in Fig.~\ref{fig: brazilian_network}) within which the nodes remain internally synchronized are well known in power engineering  and called a system split. They occur also in much larger grids, such as the European one, where such a split happened in January 2021 \cite{ENTSOEreport}, see Fig.~\ref{fig: system_split}.
We artificially increase the load (by uniformly scaling all power inputs) on the network to model a situation in which the network is operating close to capacity, with a realistic safety margin to overload ($\frac{K_{ij} - f_{ij}}{K_{ij}}\approx 10 \%$ for the most loaded line \cite{machowski_simplified_2015}, where $f_{ij}$ is the power flow from $i$ to $j$, see Sec.~\ref{subsec41}). This network provides a generic test case in the sense that approximate ratios between power input, inertia and damping are typical for power grids; see \cite{machowski_power_2008,nishikawa_comparative_2015}. It is small enough that we can computationally analyze it in detail, but large enough to provide interesting results.

As mentioned before, the power inputs are all artificially scaled to produce a grid closer to overload, to model a heavily loaded grid. The resulting set of parameters is given in Appendix A.
After swapping the values of $P_i$, $M_i$ and $\gamma_i$ for the third and fourth nodes in Fig.~\ref{fig: brazilian_network}(b) and (c), the power input is again rescaled to a safety margin of $9.1 \%$. Fluctuations are artificially added at node $4$ and node $6$ for Fig.~\ref{fig: brazilian_network}(a), at nodes $3$ and $4$ for (b), and at all nodes in (c). For (a), the line connecting node $4$ and node $5$ (with $K_{45} = 13.624$)  is the one closest to overload, as indicated in Fig.~\ref{fig: brazilian_network}(a) with a dotted line. In (b) and (c) the most heavily loaded lines are those connecting node $5$ to nodes $1$ and $2$, with $K_{15} = 4.082$ and $K_{25} = 4.444$.
For heavily loaded grids, the safety margin, $(K_{ij} - f_{ij})/K_{ij}$ for the most heavily loaded line, is generally on the order of $5 \% - 30\%$ \cite{machowski_power_2008, machowski_simplified_2015}. Note that inertia $\bm{M}$ and damping $\bm{\gamma}$ are roughly proportional to the power input at the respective nodes, but about an order of magnitude smaller than $\bm{P}$.
The calculations that we performed to determine the parameters of the Brazilian grid follow \cite{nishikawa_comparative_2015}, they are not specific to the Brazilian grid. In general, also for different grids like IEEE test-cases \cite{birchfield_grid_2017, xu_creation_2017, xu_modeling_2018}, the (order of magnitude of the) ratios between $\bm{P}$ and $\bm{M}$ and $\bm{\gamma}$ will be similar to the ones used here. What is furthermore common to power grids, including the Brazilian grid, is that most nodes have only little generation/load, and a few nodes have a very high generation or very high load, rather than power generation/load being spread relatively evenly over the grid. With increasing renewables, however, more inductive and less synchronous machines will be attached, leading to an associated decrease in the values of $M_i$ and $\gamma_i$ as compared to the total power input at that node.

\subsection{Wind versus solar data}\label{subsec23}
\begin{figure}
    \centering
    \subfloat[]{\includegraphics[width = 0.45 \textwidth]{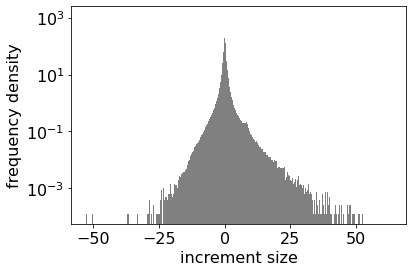}} \quad
    \subfloat[]{\includegraphics[width = 0.45 \textwidth]{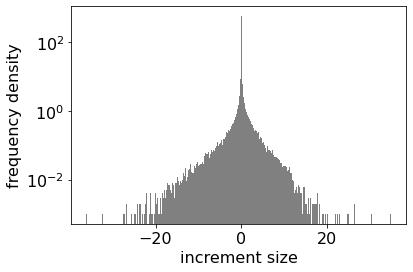}}
    \caption{Histogram of $1$-second power increments meant to be approximated by $p(t + \tau) - p(t)$, with $10^3$ bins, for (a) wind data and (b) solar data (for the brightest hour of the day) from \cite{weather_data}, showing non-Gaussian tails. Standardized for comparison (i.e. rescaled such that both figures have zero mean and unit standard deviation). These histograms should be reproduced by our data implementation.}
    \label{fig: data_histograms}
\end{figure}

Our noise terms $\xi_i$ will be constructed such that the resulting increments of power input $p_i(t +\tau) - p_i(t)$ for a given time interval $\tau$ match a realistic power input distribution. For the power input increments we will consider both Gaussian distributions and distributions taken from data on wind and solar power generation \cite{anvari_short_2016,weather_data} (representing a month of power generation in Bremen and Oldenburg, respectively), with data resolution $\tau = 1\,\text{s}$. We will vary the strength of the fluctuations by rescaling the power generation. Histograms for the distribution of power increments from \cite{weather_data} are given in Fig.~\ref{fig: data_histograms}. For  estimating desynchronization times  we consider only wind data and will approximate histogram (a) with constructed increments $p_i(t + \tau) - p_i(t)$. For a comparison of the data implementation procedure we consider also solar data. Moments are straightforward to calculate from the data, cumulants can then be calculated from the moments according to e.g. \cite{smith_recursive_1995}. The first $8$ cumulants of these distributions are shown in Table~\ref{tab: cumulants}.
\begin{table}[b]
    \caption{%
    The $n-th$ cumulants of the increments $p(t + \tau) - p(t)$ for a Gaussian distribution and for the wind and solar power generation data corresponding to Fig.~\ref{fig: data_histograms}, all standardized for comparison (i.e. rescaled to have zero mean and unit standard deviation). The values show that higher order cumulants can become quite large.}
    \begin{tabular}{| l|c c c c c c c c|}
    \hline
    Distribution / n& $1$ & $2$ & $3$ & $4$ &$5$ & $6$ & $7$ & $8$\\ \hline
    Gaussian  & $0$ & $1$ & $0$ & $0$ & $0$ & $0$ & $0$ & $0$\\
    Wind & $0$ & $1 $ & $4.198$ & $141.5$& $2.826 \cdot 10^3$ & $1.113 \cdot 10^5$ & $3.847 \cdot 10^6$ & $1.836 \cdot 10^8$ \\
    Solar & $0$ & $1 $& $-2.497$ & $124.6$& $-898.7$ & $3.992 \cdot 10^4$ & $-4.310 \cdot 10^5$ & $2.135 \cdot 10^7$\\
    \hline
    \end{tabular}
    \label{tab: cumulants}
\end{table}
The authors of \cite{tyloo_finite-time_2022} explicitly analyze the time correlations $\frac{\langle p(t) p(t') \rangle}{\langle p(t)^2 \rangle}$ for the wind and solar data of \cite{weather_data} and find that correlations typically decay on a timescale of $2-5$ minutes. Accordingly, we here set $1/\alpha = 100\,\text{s}$, and also consider $1/\alpha = 300\,\text{s}$ in Sec.~\ref{sec: results}.

\section{Data Implementation}\label{sec3}
Real data sets of power production are often presented by the corresponding distribution of power increments, usually showing significantly broader tails than Gaussian distributions. In the past, fluctuations in production and consumption were often modelled as Gaussian white noise. To overcome this approximation, we develop a new version of implementing colored noise (Sec.~\ref{subsec32}). We compare this approach with an implementation as a compound Poisson process (Sec.~\ref{subsec31}) and point to the importance of time correlations in the fluctuations.

After all, we want to approximate data in terms of fluctuating increments. 
To reproduce the measured increments, we should construct a corresponding stochastic process $\xi_i(t)$ for which the resulting distribution of $p_i(t + \tau) - p_i(t)$ accurately reproduces the distribution of data $\text{pr}(\text{increment data})$. We can either quantify $\xi_i(t)$ directly, or, equivalently, quantify it by the distribution of its increments $Z_{\Delta t,i} \equiv \int_{t}^{t+ \Delta t}\xi_i(t') \text{d}t'$ for any given $\Delta t$.

Let us start with the first option.  
The noise term $\xi_i(t)$ is in general singular, such that we cannot define an instantaneous probability distribution of $\xi_i$ \cite{kampen_stochastic_1992}. Instead we can specify $\xi_i$ by its cumulants $\Gamma^{\xi_i}_n$ \cite{kampen_stochastic_1992, smith_recursive_1995}, defined by:
\begin{alignat}{1}
    \langle \langle \xi_i(t_1)\cdot \dots \cdot \xi_i(t_n) \rangle \rangle = \Gamma^{\xi_i}_n \delta(t_1 - t_2) \cdot \dots \cdot \delta(t_1 - t_n),  \label{eq: xi_cumulants}
\end{alignat}
where we will always rescale the noise term such that $\Gamma^{\xi_i}_1 = 0$ and $\Gamma^{\xi_i}_2 = 1$. As defined in Sec.~2.2, the double angular brackets $\langle \langle \dots \rangle \rangle$ denote cumulants rather than moments, e.g. $\langle \langle x^2 \rangle \rangle = \langle x^2 \rangle - \langle x \rangle^2$. Here we will always assume that all cumulants are finite, and that the cumulant generating function $ C^{\xi_i}(J) = \sum_{n = 1}^{\infty} \Gamma^{\xi_i}_n J^n/n!$ exists for all real $J$.
For a more detailed description of white noise processes we refer to the discussions in \cite{kampen_stochastic_1992,gardiner_stochastic_2009}. The cumulants $\Gamma^{\xi_i}_n$ are constructed such that increments of the power input $p_i(t +\tau) - p_i(t)$ for a given time interval $\tau$ match a realistic power input distribution.
The cumulants $\Gamma^{\xi_i}_n$ can be related to the cumulants of $p_i(t + \tau) - p_i(t)$ by considering that Eq.~\ref{eq: p_dynamics} gives $p_i(t) = \sigma_i \int_{-\infty}^t \dd t' \, \exp(-\alpha (t - t')) \xi_i(t')$, that the cumulants of a sum are the sum of the  cumulants \cite{kampen_stochastic_1992}, and that multiplication by a constant multiplies the $n$-th cumulant by that constant to the power $n$. Using Eq.~\ref{eq: xi_cumulants} we then get that $\Gamma_n^{p_i(t + \tau) - p_i(t)}$ are related to $\Gamma^{\xi_i}_n$ according to:
\begin{alignat}{1}
    \Gamma_n^{p_i(t + \tau) - p_i(t)} &= \Gamma^{\xi_i}_n \cdot \sigma_i^n \int_{-\infty}^{t + \tau} \dd t' \, \Big[ \exp(-\alpha (t + \tau- t')) - \mathbbm{1}(t' \leq t) \exp(-\alpha (t - t')) \Big]^n \\
    &= \Gamma^{\xi_i}_n \cdot \sigma_i^n  \Big(\int_{-\infty}^{t} \dd t' \, \exp(-n\alpha (t - t')) \big[\exp(-\alpha \tau) - 1 \big]^n \\
    & \qquad + \int_{t}^{t + \tau} \dd t' \, \exp(-n\alpha (t + \tau - t')) \Big) \nonumber \\
    &= \Gamma^{\xi_i}_n \cdot \frac{\sigma_i^n }{n \alpha} \Big(\big[\exp(-\alpha \tau) - 1 \big]^n + \big[1 - \exp(- n \alpha \tau)\big] \Big) \label{eq: p_from_xi}.
\end{alignat}
Note that $\Gamma_n^{p_i(t + \tau) - p_i(t)}$ are constructed directly from the dynamics of $p(t)$ according to Eq.~\ref{eq: p_dynamics} including the parameter $\alpha$ and therefore are the quantities that should be primarily identified with the data.
To get the statistics of $p_i(t + \tau) - p_i(t)$ to approximate the power increments found from the data, i.e. $\Gamma_n^{p_i(t + \tau) - p_i(t)} \stackrel{\approx}{\rightarrow} \Gamma_n^{\text{data}}$ where $\stackrel{\approx}{\rightarrow}$ means 'should approximate', we can then reverse the relation of Eq.~\ref{eq: p_from_xi} to get that $\Gamma_n^{\xi_i}$ should be chosen such that:
\begin{alignat}{1}
    \sigma_i^n \Gamma_n^{\xi_i} &\stackrel{\approx}{\rightarrow} \Gamma_n^{\text{data}} \cdot  n\alpha  \cdot \Big(\big[\exp(-\alpha \tau) - 1 \big]^n + \big[1 - \exp(- n \alpha \tau)\big] \Big)^{-1} \,. \label{eq: xi_from_p}
\end{alignat}
The alternative characterization of $\xi_i(t')$ is by the distribution of its increments $Z_{\Delta t,i} \equiv \int_{t}^{t+ \Delta t}\xi_i(t') \text{d}t'$ for any given $\Delta t$. From here on we always consider $\Delta t$ to be measured in units of seconds, even if the units are not explicitly written out, to avoid cluttering the expressions. The cumulants of $Z_{\Delta t,i}$ can be expressed in terms of $\Gamma_n^{\xi_i}$. They are found from Eq.~\ref{eq: xi_cumulants} by using that the cumulants of a sum equal the sum of the cumulants:
\begin{alignat}{1}
    \Gamma_n^{Z_{\Delta t, i}} = \Delta t \, \Gamma_n^{\xi_i}. \label{eq: z_from_xi_cumulants}
\end{alignat}
Which of the two representations of the noise (in terms of the cumulants of  $\xi_i$ or  $Z_{\Delta t, i}$) is used for replacing $\Gamma_n^{p_i(t + \tau) - p_i(t)}$ is then a matter of convenience that depends on the application.

\subsection{Numerical implementation and generation of increments $Z_{\Delta t}$} \label{subsec32}
The swing equations are implemented numerically by Eulers method \cite{protter_euler_1997,chen_simulation_2012} with step-size $\delta t = 10^{-3}\,\text{s}$. The power input fluctuations $p_i(t)$ are updated accordingly:
\begin{alignat}{1}
    p_i(t + \delta t) - p_i(t) &=  -\alpha_i p_i(t) \cdot \delta t + \sigma_i Z_{\delta t,_i}  + \text{h.o.} \,
    \label{eq:17}\\
    Z_{\delta t,i} &= \int_t^{t + \delta t} \text{d}t' \xi_i(t'). \label{eq: xi_increment}
\end{alignat}
Euler's method corresponds to dropping the higher order terms (h.o.). At each time-step the random variables $Z_{\delta t,i}$ should be drawn from $\text{pr}\big(Z_{\delta t,i}\big) = \text{pr}\big(\int_t^{t + \delta t} \text{d}t' \xi_i(t')\big)$.
Although $\Gamma_n^{Z_{\delta t,i}} = \delta t \cdot \Gamma_n^{\xi_i}$ themselves are determined from $p_i(t + \tau) - p_i(t)$ according to Eq.~\ref{eq: xi_from_p}, numerically constructing a distribution from its cumulants is in general not tractable for strongly non-Gaussian distributions \cite{blinnikov_expansions_1998, chateau_gram-charlier_2017}. Here we will give our solution to this problem, and compare it to the alternative approach of \cite{hindes_network_2019} in Sec.~\ref{subsec31}.  We may expand the cumulants $\Gamma_n^{Z_{\Delta t, i}}$ (for $n \geq 2$) in $\alpha \Delta t$:
\begin{alignat}{1}
    \sigma^n \Gamma_n^{Z_{\Delta t,i}} &= \Gamma_n^{p_i(t + \Delta t) - p_i(t)} \cdot  n\alpha \Delta t   \cdot \Big(\big[\exp(-\alpha \Delta t) - 1 \big]^n + \big[1 - \exp(- n \alpha \Delta t)\big] \Big)^{-1} \nonumber \\
    &= \Gamma_n^{p_i(t + \Delta t) - p_i(t)} \cdot  [1 + O([\alpha \Delta t])] .\label{eq: alphatau_approx}
\end{alignat}
Since $\sigma_i^n \Gamma_n^{Z_{\Delta t,i}}$ are the cumulants of $\sigma_i Z_{\Delta t,i}$, and since $\alpha \tau \approx 10^{-2}$ is small, setting $\Delta t = \tau$ in Eq.~\ref{eq: alphatau_approx} means that the difference of using $\text{pr}(\text{increment data})$ for constructing either $\pr \big(p_i(t + \tau) - p_i(t) \big)$ or directly $\text{pr}(\sigma_i Z_{\tau,i})$ is of $O([\alpha \tau])$. Thus, neglecting the $O(\alpha \tau)$ term:
\begin{alignat}{1}
    \pr( \sigma_i Z_{\tau,i}) \approx \pr \big(p_i(t + \tau) - p_i(t) \big) \stackrel{\approx}{\rightarrow} \pr(\text{increment data}) \,, \label{eq: Z_from_data}
\end{alignat}
such that the distribution $Z_{\tau,i}$ can simply be approximated by the distribution of the increments found from the data, rescaled by $\sigma_i$. From here on we will therefore set:
\begin{alignat}{1}
    \pr(\sigma_i Z_{\tau,i}) = \pr(\text{increment data}) \,. \label{eq: data_identification}
\end{alignat}

\begin{figure}
    \centering
    \hspace{0.2cm}
    \begin{tabular}{ll}
    \subfloat[]{\includegraphics[width = 0.45 \textwidth]{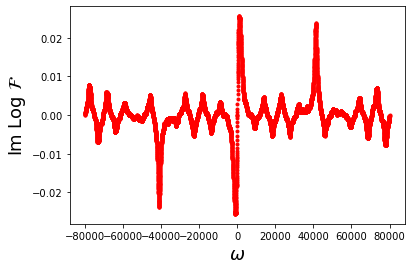}}
    &
    \subfloat[]{\includegraphics[width = 0.45 \textwidth]{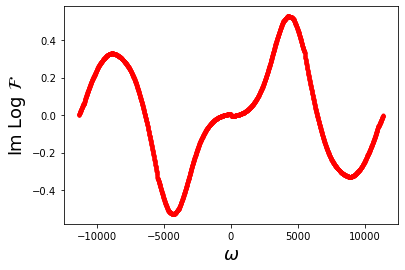}}
    \end{tabular}
    \caption{$\text{Im} \, \log \big( \mathcal{F} (\pr(Z_\tau))[w] \big)$ for the standardized wind data (a) and solar data (b) from \cite{weather_data}, discretized in $10^4$ and $3 \cdot 10^3$ bins respectively. The values remain in $(- \pi, \pi]$, such that within the considered range of frequencies the principal branch of the logarithm satisfies the requirement that the Fourier transform is continuous.}
    \label{fig: fourier}
\end{figure}

\begin{figure}
    \centering
    \hspace{0.2cm}
    \begin{tabular}{lll}
    \subfloat[]{\includegraphics[width = 0.32 \textwidth]{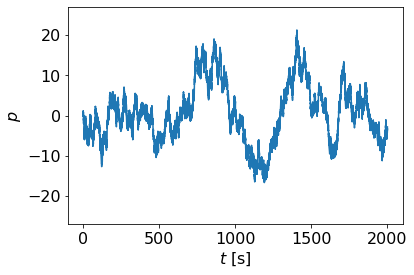}}
    &
    \subfloat[]{\includegraphics[width = 0.32 \textwidth]{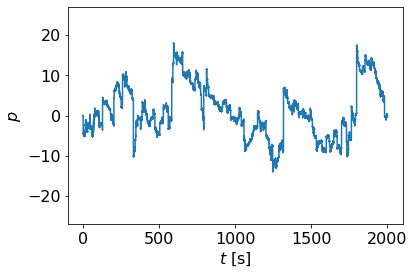}}
    &
    \subfloat[]{\includegraphics[width = 0.32 \textwidth]{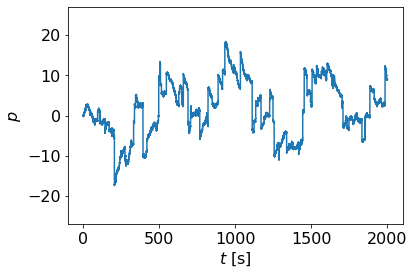}}
    \end{tabular}
    \caption{Example of time-series generated from Eq.~\ref{eq: p_dynamics} by the procedure of Sec.~\ref{subsec32} with $\alpha = 10^{-2}\,\text{s}$, for various (standardized) distributions of $p(t + \tau) - p(t)$. a) Gaussian distribution. b) Wind power generation \cite{weather_data}. c) Solar power generation during the sunniest hour of the day \cite{weather_data}.}
    \label{fig: p_time_series}
\end{figure}

What is left is using this distribution of $Z_{\tau,i}$ to construct the distribution of $Z_{\delta t,i}$ for the finer time resolution $\delta t$. \footnote{Formally, random variables $Z_\tau$ are only increments of a (time-homogeneous) white noise process if the distribution of $Z_\tau$ is \textit{infinitely divisible}; see \cite{steutel_infinite_2004}. Here we assume this is the case (to a sufficiently good approximation). If $\pr(Z_\tau$) is not infinitely divisible, then for $\delta t \rightarrow 0$ the distribution of $Z_{\delta t}$ calculated by the procedure shown in this section may give complex probabilities rather than real, positive numbers. We have not found this to be the case for $\delta t = 10^{-3} - 10^{-4}\,\text{s}$ and discretization up to several thousand bins.
} Using the fact that the Fourier transform of the distribution of a sum of independent random variables equals the product of the Fourier transforms of the distributions of the random variables \cite{kampen_stochastic_1992}, we get:
\begin{alignat}{1}
    \mathcal{F}\big(\pr(Z_\tau)\big)[\omega] &= \Big[\mathcal{F}\big(\pr(Z_{\delta t})\big)[\omega]\Big]^{\tau/\delta t} \,.\label{eq:fourier}
\end{alignat}
For Gaussian $\text{pr}(Z_\tau) = N(0,1)$, this is solved by $\pr(Z_{\delta t}) = N(0, \delta t/\tau)$, in which case (for $\text{pr}(Z_\tau)$ given) the standard Euler method for Gaussian SDEs is obtained \cite{gardiner_stochastic_2009}. In general, Eq.~\ref{eq:fourier} should be solved for $\pr(Z_{\delta t})$ numerically. \\
Once $\pr(Z_\tau)$ is discretized (i.e. represented as a histogram), $\mathcal{F}\big(\pr(Z_\tau)\big)[\omega]$ is easily found by a discrete Fourier transform (for which we used the Fast Fourier Transform library implemented in Scipy \cite{2020SciPy-NMeth}). Similarly, once $\mathcal{F}\big(\pr(Z_{\delta t})\big)[\omega]$ is found, $\pr(Z_{\delta t})$ is easily obtained by the inverse discrete Fourier transform. The complex logarithm is multi-valued, however, such that:
\begin{alignat}{1}
    \mathcal{F}\big(\pr(Z_{\delta t})\big)[\omega] &=  \exp\Big(\frac{\delta t}{\tau} \Big[\Log \mathcal{F}\big(\pr(Z_\tau)\big)[\omega] + n(\omega) \cdot 2 \pi i \Big] \Big), \quad n(\omega) \in \mathbb{Z} \label{eq:24}
\end{alignat}
where $\Log$ denotes the principal branch of the logarithm (i.e. with imaginary part in $(- \pi, \pi]$).
The Fourier transform $\mathcal{F}\big(\pr(Z_{\delta t})\big)[\omega]$ is continuous in $\omega$ and equals zero for $\omega = 0$, so $n(\omega)$ should be chosen accordingly. Discretizing $\text{pr}(Z_\tau)$ with resolution $\Delta Z_\tau$ requires frequencies up to $\omega = \pm \frac{\pi}{\Delta Z_\tau}$; even when discretizing the $\text{pr}(Z_\tau)$ taken from the data by up to $10^4$ bins, $\text{Im} \Big(\Log \mathcal{F}\big(\pr(Z_\tau)\big)[\omega] + n(\omega) \cdot 2 \pi i \Big) \in (- \pi, \pi]$, such that we can simply set $n(\omega) = 0$ (see Fig.~\ref{fig: fourier})
\footnote{$\text{Im} \, \log \big( \mathcal{F} (\pr(Z_{\delta t}))[w] \big)$ can be shown to be equal to $\frac{\delta t}{\tau} \times \big(\Gamma^{Z_\tau}_1 \, \omega - \Gamma^{Z_\tau}_3 \, \omega^3/3! + \Gamma^{Z_\tau}_5 \, \omega^5/5! \dots \big)$. A translation of the distribution $\text{pr}(Z_\tau)$ along the $x$-axis changes $\Gamma^{Z_\tau}_1$ and hence changes the linear term of $\text{Im} \, \big(\mathcal{F} (\pr(Z_{\delta t}))[w]\big)$. We found that in order for setting $n(\omega) = 0$ to be valid it is important to ensure that the mode of the discretized distribution $\pr(Z_\tau)$ equals $0$, by shifting the $x$-axis if this is not the case.}.
Once $\pr(Z_{\delta t,i})$ is found, random increments $Z_{\delta t, i}$ can be generated with standard sampling methods; we used 'rv\_histogram' implemented in Scipy \cite{2020SciPy-NMeth}. Examples of the resulting time-series are shown in Fig.~\ref{fig: p_time_series} for different distributions. We term our method, based on Eq.~\ref{eq:fourier}, the Fourier-method in the following.
\subsection{Approximation by a compound Poisson process and its small $\tau$-approximation}\label{subsec31}
An alternative method for reconstructing non-Gaussian noise from data was used in \cite{hindes_network_2019}, without explicitly mentioning the involved approximations, by representing the noise by a compound Poisson process.
A continuous time Markov process, in this case assumed for $p(t)$, can be characterized by the rescaled moments of its transition distribution, the Kramers-Moyal coefficients $M^{(n)}(p)  \equiv \lim_{\Delta t \rightarrow 0}\frac{1}{\Delta t}\langle \big(p(t + \Delta t) - p(t) \big)^n| p(t) = p \rangle$ \cite{gardiner_stochastic_2009, anvari_disentangling_2016}. The procedure of \cite{hindes_network_2019} effectively approximates the Kramers-Moyal coefficients for $n \geq 2$ (which, for the process of Eq.~\ref{eq: p_dynamics}, are independent of $p$) by their values for finite $\Delta t$, with $\Delta t = \tau$ equal to the data resolution. The procedure of \cite{hindes_network_2019} approximates the correct Kramers-Moyal coefficients arbitrarily well for data resolution $\tau \rightarrow 0$ (see \cite{anvari_disentangling_2016}, or the derivations that follow in this section). 
Here we will show, taking Eq.~\ref{eq: alphatau_approx} as a starting point, that this compound Poisson ('CP') approximation is an approximation in addition to that of Eq.~\ref{eq: alphatau_approx} (i.e. the approximation entering also the Fourier-method), and explicitly derive the error terms for the CP-approximation. It turns out that for finite $\Delta t$ the Kramers-Moyal coefficients are better approximated by cumulants of the data than by its moments; the Fourier-Method corresponds to the former, while the CP approximation corresponds to the latter.
{} \\
We will find that although the procedure of \cite{hindes_network_2019} is justified for non-Gaussian noise in the limit of infinite measurement resolution $\tau \rightarrow 0$, it is uncontrolled in the sense that it is not clear $\grave{a}$ priori how small $\tau$ should be for this mapping to be a good approximation (see also \cite{lehnertz_characterizing_2018}). For $\tau = 1\,\text{s}$, the approximation is relatively accurate for the wind generation data from \cite{weather_data}, while it gives significant errors for the solar generation data from \cite{weather_data}, as we discuss below in Table 2, and fails for Gaussian fluctuations (for which the authors of \cite{hindes_network_2019} do not use the compound Poisson process), in contrast to our Fourier-method that applies also to Gaussian noise.
{} \\
Under mild assumptions \cite{gardiner_stochastic_2009}, the formal time-derivative of any continuous-time Markov process, and in particular $\xi_i(t)$ (entering in $\dot{p}_i(t)$ according to Eq.~\ref{eq: p_dynamics}), can be decomposed into a sum of Gaussian noise and a compound Poisson process; see \cite{gardiner_stochastic_2009,steutel_infinite_2004,anvari_disentangling_2016} for derivations \footnote{Gaussian noise  itself can be obtained as the limit of a sequence of compound Poisson processes \cite{kampen_stochastic_1992,steutel_infinite_2004}, but this will not be useful for our purposes.}. The Gaussian noise generates continuous Brownian motion of $p_i(t)$, and the compound Poisson process generates discontinuous 'jumps'. A compound Poisson process consists of a series of delta functions, randomly distributed over time with the average number of pulses per second given by a rate $\rho$. Each of the delta pulses has a prefactor $z$ randomly drawn from a size distribution $\lambda(z)$, which determines the distribution of jump sizes of $p_i(t)$. If the set of possible jump sizes $z$ is discrete, the compound Poisson process may alternatively be seen as a sum of Poisson processes, each with its own rate and its own jump size $z$ \cite{hindes_network_2019, kampen_stochastic_1992}, which makes it 'compound'.

The presence of jumps can be visualized in Fig.~\ref{fig: p_time_series}: For Fig.~\ref{fig: p_time_series}(a) the noise $\xi_i$ is Gaussian, and $p_i(t)$ follows a continuous (but non-differentiable \footnote{That is, the formal derivative $\dot{p}_i(t)$ does not have a finite value for any $t$ \cite{kampen_stochastic_1992,gardiner_stochastic_2009}.}) path, while for non-Gaussian noise (Figs.~\ref{fig: p_time_series}(b-c) this continuous movement is randomly interrupted by sudden 'jumps' with various jump sizes. Discontinuous jumps in $p_i(t)$ correspond to delta functions in its derivative, i.e. in $\xi_i(t)$. If the contribution of the Brownian motion part of $p_i(t)$ is assumed to be negligible compared to that of the discontinuous jumps,  we may approximate $\xi_i(t)$ by a compound Poisson process.
{} \\
For a given time interval $\Delta t$ (not necessarily equal to the discrete time-steps in the numerical simulation), the random increments $Z^{\text{cp}}_{\Delta t}\equiv \int_t^{t + \Delta t} \dd t' \xi^{\text{cp}}(t')$ of a compound Poisson process $\xi^{\text{cp}}(t)$ can be generated as follows \cite{kampen_stochastic_1992}:
\begin{itemize}
    \item The total number of pulses in the time interval $[t, t + \Delta t)$ is generated as a Poisson random variable $x_{\Delta t}$ with parameter $\rho \Delta t$.
    \item The contribution of each of the pulses is assigned a random size $z$, independently drawn from a distribution $\lambda(z)$. The increment $Z^{\text{cp}}_{\Delta t}$ is then the sum of the sizes $z$ of the contribution of each of the pulses.
\end{itemize}
Analytically, instead of subsequently generating the compound Poisson process, an alternative representation of $Z^{\text{cp}}_{\Delta t}$ is more convenient:  we can discretize $z$ according to $z \in \{\dots, -\Delta z, 0, \Delta z, \dots \}$ with $\Delta z \rightarrow 0$ and denote the number of pulses in $[t, t + \Delta t)$ with contribution smaller than $z$ by $x_{\Delta t}(z)$. Then the increments can be written as $Z^{\text{cp}}_{\Delta t} =  \sum_z z \cdot \Delta  x_{\Delta t}(z)$, where $\Delta x_{\Delta t}(z) \equiv x_{\Delta t}(z + \Delta z) - x_{\Delta t}(z ) $ give the number of pulses with size in $[z, z + \Delta z)$, which are independent Poisson random variables with parameter $\rho  \Delta t \lambda(z) \cdot \Delta z$. 
If the statistics of the compound Poisson process $\xi^{\text{cp}}(t)$ should approximate those of $\xi(t)$, $\rho$ and $\lambda(z)$ should be chosen such that the increments $Z^{\text{cp}}_{\Delta t}$ have the same distribution as $Z_{\Delta t,i} \equiv \int_t^{t+\Delta t} \dd t' \xi_i(t')$. The random increments $Z^{\text{cp}}_{\Delta t}$ have the cumulant generating function:
\begin{alignat}{1}
    \log \langle \exp(J \cdot Z^{\text{cp}}_{\Delta t} ) \rangle_{\pr(Z^{\text{cp}}_{\Delta t})} &= \sum_z \log \langle \exp(J \cdot z \cdot\Delta  x_{\Delta t}(z) ) \rangle_{\pr(\Delta  x_{\Delta t}(z))}  \nonumber\\
    &= \sum_z \log \Big(\sum_{n = 0}^{\infty} \exp(J z n) \pr(\Delta  x_{\Delta t}(z) = n)\Big) \nonumber \\
    &= \sum_z \log \Big(\exp\big(- \rho \Delta t \lambda(z) \Delta z\big) \cdot \sum_{n = 0}^{\infty} \frac{\big[\exp(J z) \rho \Delta t \lambda(z) \Delta z \big]^n}{n!}\Big) \nonumber\\
    &= \rho {\Delta t} \cdot \sum_z \lambda(z) (\exp(J \cdot z) - 1) \Delta z \nonumber  \\
    &= \rho {\Delta t} \cdot \big(\langle \exp(J \cdot z) \rangle_{\lambda(z)} - 1 \big)\,,
\end{alignat}
where in the first line we used that the cumulant generating function of a sum is the sum of the cumulant generating functions of the individual variables \cite{kampen_stochastic_1992,gardiner_stochastic_2009}, and in the second to fourth lines we evaluated the expectation value over $\Delta x_{\Delta t}(z) \sim \text{Poiss}\big(\rho \Delta t \lambda(z) \Delta z\big)$.  \\
The left-hand side is the cumulant generating function of $\text{pr}(Z^{\text{cp}}_{\Delta t})$, while the right-hand side is (up to a constant) proportional to the moment generating function of $\lambda(z)$. Thus:
\begin{alignat}{1}
    \rho \Delta t \mu^{\lambda}_n = \Gamma^{Z^{\text{cp}}_{\Delta t}}_n = \Delta t \Gamma^{\xi^{\text{cp}}}_n \,, \label{eq: poiss_moments}
\end{alignat}
where $\mu^{\lambda}_n$ are the moments of $\lambda(z)$. To construct the compound Poisson process such that the random variables $Z^{\text{cp}}_{\tau}$ follow approximately the same distribution as $Z_{\tau,i}$, we find from setting $\Delta t = \tau$ in Eq.~\ref{eq: poiss_moments} that $\rho$ and $\lambda(z)$ should be such that:
\begin{alignat}{1}
    \rho \mu^{\lambda}_n &\stackrel{\approx}{\rightarrow}  \Gamma^{Z_{\tau ,i}}_n/\tau  \,. \label{eq: approx_poiss_moments}
\end{alignat}
Now it would seem like we have not come any closer to approximating $\xi$, since in general finding a distribution $\lambda(z)$ with these moments is not straightforward \cite{blinnikov_expansions_1998, chateau_gram-charlier_2017}. However, for small $\tau$ the cumulants and moments of $Z_{\tau,i}$ approach each other:
\begin{alignat}{1}
    \langle \exp(J \cdot Z_{\tau,i}) \rangle_{\pr(Z_{\tau_i})} - 1\, &=  \exp \log \big( \langle \exp(J \cdot Z_{\tau,i}) \rangle_{\pr(Z_{\tau,i})} \big) - 1\, \nonumber \\
    &=  \log \langle \exp(J \cdot Z_{\tau,i} ) \rangle_{\pr(Z_{\tau,i})} + O\big( \big[\log \langle \exp(J \cdot Z_{\tau,i} ) \rangle_{\pr(Z_{\tau,i})}\big]^2 \big) \nonumber \\
    &=  \log \langle \exp(J \cdot Z_{\tau,i} ) \rangle_{\pr(Z_{\tau,i})} + O\big(\tau^2) \,, \label{eq: small_tau}
\end{alignat}
where we expanded the exponential in its Taylor series and used that by Eq.~\ref{eq: z_from_xi_cumulants} the cumulant generating function $\log \langle \exp(J \cdot Z_{\tau,i} ) \rangle_{\pr(Z_{\tau,i})} = \Gamma^{Z_{\tau_i}}_1 \frac{J}{1!} + \Gamma^{Z_{\tau_i}}_2 \frac{J^2}{2!} + \dots\sim \tau$. If the distribution of $Z_{\tau,i}$ is neither Gaussian nor a delta function, all of its cumulants are non-zero \cite{gardiner_stochastic_2009}. Expanding both the moment generating function (the left-hand side of Eq.~\ref{eq: small_tau}) and the cumulant generating function (on the right-hand side of Eq.~\ref{eq: small_tau}) in $J$, we then find that for all orders $\mu_n^{Z_{\tau,i}} = \Gamma_n^{Z_{\tau,i}}[1 + O(\tau)]$. Neglecting the $O\big(\tau)$ term, Eq.~\ref{eq: approx_poiss_moments} is then solved by setting $\rho = 1/\tau$ and $\lambda(z) = \pr(Z_{\tau,i} = z)$. Therefore, consistently, we may replace $\mu^{\lambda}_n$ with $\mu_n^{\text{data}}$ read off from the data. After all, this means that $\Gamma^{\xi^{\text{cp}}}_n$ can be replaced by $\mu_n^{\text{data}}$ which we will use in a comparison between both methods in Sec.~\ref{sec: results}. \\

The identification $\lambda(z) = \pr(Z_{\tau,i} = z)$ may be interpreted as follows: $\xi_i$ can always be composed into the sum of Gaussian noise and a compound Poisson process with a certain ($\tau$-independent) rate. If for small $\tau$ the increments $Z_{\tau,i}$ are dominated by the contribution of the compound Poisson process, then in the limit $\tau \rightarrow 0$ the probability of having two or more pulses in the time interval $[t, t + \tau)$ becomes negligible, such that the distribution of pulse contributions $\lambda(z)$ becomes (up to an irrelevant delta function at $z= 0$ corresponding to having no pulses at all) equal to the increment distribution $\pr(Z_{\tau,i} = z)$.

The case of $\xi_i$ being Gaussian noise is an exception: In Eq.~\ref{eq: small_tau} the leading order contribution to $\mu_n^{Z_{\tau,i}}$ is the $O\big(\tau^2)$ term (all cumulants of order $m > 2$ are zero), which can no longer be neglected. While for the non-Gaussian case the approximation is valid for $\tau \rightarrow 0$, it is not obvious how small $\tau$ should be for the approximation to be accurate. \footnote{This is in contrast with the approximation of Eq.~\ref{eq: alphatau_approx}, where $\alpha \tau$ is unitless, and the approximation (of replacing $\text{pr}(\sigma_i Z_{\tau,i})$ by $\text{pr}(p_i(t+\tau)-p_i(t))$) can thus be seen to be justified whenever $\tau \ll 1/\alpha$. }

\section{Dimensional Reduction of the equations of motion}\label{sec4}
The different versions of dimensional reduction of the equations of motion given in this section are based on the fact that an instability of the synchronized state is caused by overloaded lines. To apply these reductions, we first have to identify transmission lines that are endangered by possible overloads (Sec. 4.1). In Sec. 4.2 we apply the so-called synchronized subgraph approximation to the swing equations (Eqs.1-2,3), for which desynchronization times are calculated by stochastic sampling with many iterations. As an alternative analytical approach, applicable to very long desynchronization times, we apply the WKB-approach to the differential Chapman-Kolmogorov equation that is presented in Sec. 4.3 and derived in Appendix B. Inserting the WKB-ansatz into this equation, we derive in Sec. 4.4 Hamilton's equations of motion for the full system, including reductions to 12- and two-dimensional phase spaces, respectively. For the 12-dimensional case we later make use of the iterative action minimization method (Sec. 5.2) \cite{lindley_iterative_2013}, for the two-dimensional case we derive the analytical form of the action. The WKB-approach avoids the need for stochastic sampling in the regime where desynchronization events are very rare.

\subsection{Partition of the network: Identifying endangered transmission lines}\label{subsec41}
In the absence of fluctuations $p_i(t)$, fixed points of the swing equations satisfy (besides $v_i = 0$) for all $i$:
\begin{equation}
    P_i + \sum_j K_{ij} \sin(\phi_j - \phi_i) = 0 \,,
\end{equation}
Deterministically, the system is in stable operation as long as it resides at a fixed point which is linearly stable. Linear stability of a fixed point $\bm{\phi}^\ast$ is determined by the eigenvalues of the Laplacian matrix $L(\bm{\phi}^\ast)$ \cite{manik_supply_2014} with elements:
\begin{equation}
    L_{ik}(\bm{\phi}^\ast) \equiv \frac{\partial \sum_k K_{ik} \sin(\phi_k - \phi_i)}{\partial \phi_j} \biggr \rvert_{\bm{\phi}^\ast}\;.
\end{equation}
The Laplacian always has a zero eigenvalue corresponding to a global translation of the phase angles. Assuming the network has a single connected component, stability of the fixed point is then determined by the remaining eigenvalues. The fixed point is stable if its second highest eigenvalue (with zero being the highest eigenvalue), the Fiedler value (with the corresponding eigenvector called the Fiedler vector), is negative. The authors of \cite{manik_supply_2014} show that a sufficient (but not necessary) condition for linear stability of the fixed point is $|\phi^\ast_i - \phi^\ast_j| < \pi/2$ for any two nodes $i$ and $j$ connected by a line. In power engineering this condition is used as a standard criterion for stable synchronized operation \cite{machowski_power_2008,ENTSOEreport}, and it is this type of fixed points that we will focus on in this work.
Linear stability can then be interpreted as the lines not being overloaded:
\begin{equation}
    |\phi_i^\ast - \phi_j^\ast| < \pi/2 \quad  \leftrightarrow \quad | f_{ij} | < K_{ij} \,,\label{eq: stability_requirement}
\end{equation}
where $f_{ij} = K_{ij}\sin(\phi_i^\ast - \phi_j^\ast)$ is the power flow from node $i$ to node $j$. If the load on some parts of the grid is increased to the point where $|\phi^\ast_i - \phi^\ast_j| \rightarrow \pi/2$ (i.e., $|f_{ij}| \rightarrow K_{ij}$) for some lines $(ij)$, the system undergoes a saddle-node bifurcation, in which the stable fixed point collides with a saddle-node (for which $|\phi^s_i - \phi^s_j| > \pi/2$) and annihilates \cite{manik_supply_2014}.  In \cite{manik_supply_2014} it is shown that for any bifurcation due to an overload, the system splits into two areas (which we will call $A_1 \equiv \{i \text{ in area } 1 \}$ and $
A_2 \equiv \{j \text{ in area } 2 \}$), for which all lines connecting the two areas are overloaded: $| f_{ij} | = K_{ij}$ for all lines $(ij)$ s.t. $i \in A_1$ and $j \in A_2$. At the bifurcation, the two areas desynchronize, while the nodes remain synchronized within each area; in particular, the Fiedler vector becomes $(\sqrt{\frac{|A_2|/|A_1|}{|A_1| + |A_2|}}: i \in A_1, - \sqrt{\frac{|A_1|/|A_2|}{|A_1| + |A_2|}}: j \in A_2)$.

\subsection{Swing equations in the synchronized subgraph approximation}\label{subsec42}
Although the stable fixed point $\bm{\phi}^\ast$ is linearly stable against fluctuations $\bm{p}(t)$, strong fluctuations may still kick the system out of the basin of attraction of $\bm{\phi}^\ast$. 
As mentioned in Sec.~\ref{subsec41}, at overload the network splits into two areas, each of which remains internally synchronized. A natural reduction is then to model each of the two areas as a single oscillator, and approximate the rate of desynchronization between the two areas by the rate of desynchronization between the two reduced oscillators. Such a reduction was formulated at the level of Hamilton's equations (see Sec.~\ref{sec: wkb}) in \cite{hindes_network_2019} and dubbed the 'synchronized subgraph approximation'. Here  instead we formulate this reduction more generally, and directly at the level of the swing equations: We specify the two areas of the grid which are at risk for a system split (shown for the Brazilian grid in Fig.~\ref{fig: brazilian_network}), choose area $1$ to be the area in which the total power production $\sum_{i \in A_1} P_i = -\sum_{j \in A_2} P_j$ is positive (where the sets $A_1$ and $A_2$ are as in Sec.~\ref{subsec41}); then we write $\phi_1 = \phi_i - (\phi_i^{SN} - \frac{\pi}{4})$ for all $i \in A_1$ and $\phi_2 = \phi_j - (\phi_j^{SN} + \frac{\pi}{4})$ for all $j \in A_2$ (where $\bm{\phi}^{SN}$ are the phase angles at bifurcation), and sum the swing equations over all $i \in A_1$ and $j \in A_2$, respectively, to obtain:
\begin{alignat}{1}
    M_1 \dot{v}_1 + \gamma_1 v_1 &=  p_1(t) + P + K \sin(\phi_2 - \phi_1)  \label{eq: reduced_eq_1}\\
    M_2 \dot{v}_2 + \gamma_2 v_2 &= p_2(t) - P + K \sin(\phi_1 - \phi_2) \label{eq: reduced_eq_2} \\
    \dot{p}_1 &= - \alpha p_1 + \sigma_1 \cdot \xi_1 \label{eq: reduced_eq_3}\\
    \dot{p}_2 &= - \alpha p_2 + \sigma_2 \cdot \xi_2 \,, \label{eq: reduced_eq_4}
\end{alignat}
where:
\begin{alignat}{3}
    P &\equiv \sum_{i \in A_1} P_i = - \sum_{j \in A_1} P_j \qquad   K &\equiv \sum_{i \in A_1, j \in A_2} K_{ij}\qquad \gamma_k &\equiv \sum_{i \in A_k} \gamma_i\nonumber\\
        M_k &\equiv \sum_{i \in A_k} M_i \qquad \qquad &\sigma_k \xi_k \equiv \sum_{i \in A_k} \sigma_i \xi_i \qquad
    \sigma_k^2 &\equiv \sum_{i \in A_k} \sigma_i^2 \label{eq: xi_synchronized}
\end{alignat}
with $k\in\{1,2\}$. This thus gives a reduction of the swing equations in which each of the two areas involved in a system split is represented by a single oscillator. This reduced representation is a generalization of the reduction of \cite{hindes_network_2019} in the sense that it applies to any network topology (as long as the bifurcation is due to an overload), and is written directly in terms of the swing equations. Furthermore, our description of the noise terms in terms of their cumulants is essential for quantifying the noise terms within this dimensional reduction. The fact that the cumulants of a sum are the sums of the cumulants gives:
\begin{alignat}{1}
    \Gamma_n^{\xi_k} &= \sum_{i \in A_k} \frac{\sigma^n_i}{\sigma_k^n} \Gamma_n^{\xi_i} \qquad \quad C^{\xi_k}(\sigma_k J ) =  \sum_{i \in A_k} C^{\xi_i}(\sigma_i J ) \label{eq: cumulants_ssa}
\end{alignat}
for $k \in \{1,2\}$.
~\\
Lastly, note that the power flow between the two nodes equals $P$. The bifurcation therefore occurs at $P = K$. The stable and saddle fixed points are found to be (up to a global phase shift):
\begin{alignat}{2}
    \phi^\ast_1 &= - \phi_2^\ast =& \frac{1}{2}\arcsin(\frac{P}{K})\,, \label{eq: phi_ast}\\
    \phi^s_1 &= - \phi_2^s =& \frac{\pi}{2} - \frac{1}{2}\arcsin(\frac{P}{K})\,, \label{eq: phi_s}
\end{alignat}
respectively.

\subsection{The WKB approach} \label{sec: wkb}
To some extent, the problem becomes  analytically tractable in the limit in which typical fluctuations are small compared to the distance to overload ($\sigma \ll K - P$). To this end, we first  have to derive an equation for the time evolution of the probability to find the system in a state with variables $\bm{\phi}, \bm{v}, \bm{p}$ at time $ t$, subject to the swings equations (Eqs.~\ref{eq: swing_dynamics}-\ref{eq: p_dynamics}) and for specified initial conditions.

\subsubsection{The differential Chapman-Kolmogorov equation}\label{subsubsec43}
Under mild assumptions, further specified in \cite{gardiner_stochastic_2009}, for any continuous time Markov process, here for the swing equations with fluctuations according to Eq.~\ref{eq: p_dynamics}, the time evolution of  the probability distribution $\pr(\bm{\phi}, \bm{v}, \bm{p}, t)$ to find phases $\bm{\phi}$, frequencies $\bm{v}$, power input $\bm{p}$, at time $t$
is given by an integro-differential equation, called the differential Chapman-Kolmogorov equation \cite{gardiner_stochastic_2009}. In Appendix B we derive this evolution equation from Eqs.~\ref{eq: swing_dynamics}-\ref{eq: p_dynamics} to be given as:
\begin{alignat}{2}
    \frac{\partial \pr(\bm{\phi}, \bm{v}, \bm{p}, t)}{\partial t}
    = \sum_i \Big[ &- \frac{\partial}{\partial \phi_i}[v_i \cdot \pr(\bm{\phi}, \bm{v}, \bm{p}, t)] \nonumber \\
    &- \frac{\partial}{\partial v_i}\big[\frac{1}{M_i}\big(- \gamma_i v_i + \overline{P}_i + p_i + \sum_j K_{ij} \sin(\phi_j - \phi_i)\big) \cdot \pr(\bm{\phi}, \bm{v}, \bm{p}, t) \big]  \nonumber\\
    &- \frac{\partial}{\partial p_i} \big[-\alpha_i p_i \cdot \pr(\bm{\phi}, \bm{v}, \bm{p}, t)\big] \nonumber \\
    &+ \int \dd z_i w_i(z_i) \cdot \big(\pr(\bm{\phi}, \bm{v}, \bm{p} - \bm{\mathbbm{1}}_i \sigma_i z_i, t) - \pr(\bm{\phi}, \bm{v}, \bm{p}, t) \big) \Big] \,, \label{eq: m_eq}
\end{alignat}
with the transition rates $w_i(z_i)$ (uniquely) specified by their moment-generating function $\langle \exp(J z_i) \rangle_{w_i(z_i)} = 1 + C^{\xi_i}(J) \equiv  1 + \Gamma_1^{\xi_i} \frac{J}{1!} + \Gamma_2^{\xi_i} \frac{J^2}{2!} + \dots$. In combination with the WKB ansatz, $w_i(z)$ leads to the occurrence of the cumulant generating function $C^{\xi_i}$ in Hamilton's equation of motion below.
For a data implementation according to the Fourier-method, the cumulants are given by $\Gamma_n^{\xi_i} = \Gamma_n^{\text{data}}/(\tau \sigma_i^n) $ (Eqs.~\ref{eq: z_from_xi_cumulants}, \ref{eq: data_identification}), and $w_i(z_i)$ can in principle be numerically calculated from its Fourier transform, see Eq.~\ref{eq: w_from_fourier}. For Gaussian noise, $w_i(z_i) = \delta(z_i) + \frac{1}{2}\frac{\text{d}^2\delta(z_i)}{\text{d}z_i^2}$ turns the differential Chapman-Kolmogorov equation into a Fokker-Planck equation. If the power input is generated as a compound Poisson process, $w_i(z_i) = \rho \lambda_i(z_i) + (1 - \rho) \delta(z_i)$. In general, $w_i(z_i)$ is a weighted sum of the expression for Gaussian noise and the expression for the compound Poisson process \cite{gardiner_stochastic_2009}. For further details see Appendix B.

\subsubsection{Hamilton's equations for the full system}\label{subsubsec44}
If typical fluctuations are small compared to the distance to overload ($\sigma \ll K - P$), fluctuations that lead to desynchronization become very rare, and the rate of escape can be found by the WKB-approximation. The WKB-approximation assumes that escape from the basin of attraction of the stable fixed point becomes dominated by a single trajectory, along which escape is overwhelmingly more likely than along any other trajectory. This 'optimal path' is found by solving Hamilton's equations \cite{hindes_network_2019, dykman1, touchette_large_2009,assaf_wkb_2017} subject to boundary conditions. Initial values are at the stable fixed point, while the final values are at the edge of the basin of attraction, in this case we fix it at a saddle point.

The question remains which saddle point to choose: The system undergoes a saddle-node bifurcation whenever the lines between any two areas of the grid are overloaded. Once the lines that are in danger of being overloaded are specified, we choose as final values the saddle-node which emerges along with the stable fixed point at this bifurcation. Just like the synchronized subgraph approximation, the WKB method thus requires the lines which are in danger of being overloaded to be specified. Here we choose the partition as in Fig.~\ref{fig: brazilian_network}. In general the saddle-node which gives the lowest average time to desynchronization should be chosen.
Following \cite{dykman1, assaf_wkb_2017} we assume that in this regime the quasi-stationary distribution follows the WKB-form:
\begin{alignat}{1}
    \pr(\bm{\phi}, \bm{v}, \bm{p}, t) &\propto \exp( - S(\bm{\phi}, \bm{v}, \bm{p}, t)) \,,\label{eq:former48}
\end{alignat}
with $S \rightarrow \infty$ as the noise strength goes to zero. Often a large factor in front of $S$ in Eq.~\ref{eq:former48} is explicitly split off. The WKB-approach applies to fluctuations which are rare. For example, when WKB is  applied to large populations in the context of evolutionary game theory \cite{dykman1, assaf_wkb_2017, serrao_rare_2021},  large fluctuations are suppressed if the system size $N$ is large. Typical fluctuations in population sizes are of $O(\sqrt{N})$, rare ones are of $O(N)$. In this case a factor $N$ is split off.

For our case, if the strength of noise is parameterized by $\sigma \ll 1$, typical fluctuations in $\bm{\phi}, \bm{v}, \bm{p}$ are of $O(\sigma)$, while the WKB approach applies to rare ones of $O(1)$. For Gaussian noise, the WKB-approach is formally derived in the limit $1/\sigma^2 \rightarrow \infty$ in \cite{freidlin_random_2012,dembo_large_2010}, for which the action is proportional to $1/\sigma^2$, which may in principle be split off from the action. For the non-Gaussian case the precise dependence of $S$ on $\sigma$ is less obvious. Furthermore, in our case desynchronization events are rare either because the noise strength is too small to overcome the distance to the boundary of the basin of attraction of the synchronized state, or because the load of the system is so small that the distance to the boundary of the basin of attraction is very large, even compared to a larger noise strength.
Inserting the WKB ansatz into the differential Chapman-Kolmogorov equation, and defining auxiliary momenta $\bm{\lambda^\phi} \equiv \frac{\partial S}{\partial {\bm{\phi}}}$ , $\bm{\lambda^v} \equiv \frac{\partial S}{\partial {\bm{v}}}$ , $\bm{\lambda^p} \equiv \frac{\partial S}{\partial {\bm{p}}}$, gives:
\begin{alignat}{1}
    - \frac{\partial S}{\partial t} = \sum_i &\Big[v_i \cdot \lambda^\phi_i  - \alpha_i p_i \cdot \lambda^p_i +  \frac{1}{M_i}\big(- \gamma_i v_i + P_i + p_i + \sum_j K_{ij} \sin(\phi_j - \phi_i)\big) \cdot \lambda^v_i \Big]  \\
    & + \int \text{d}z_i w_i(z_i) \cdot \Big[\exp\big(- S(\bm{\phi},\bm{v}, \bm{p} - \mathbbm{1}_i \sigma_i z_i, t) + S(\bm{\phi},\bm{v}, \bm{p}) \big) - 1\Big] + \text{constant} \,, \label{eq:formerB8}
\end{alignat}
where we are free to set the constant to zero by absorbing it in a $t$-dependent prefactor in Eq.~\ref{eq:former48}. Expanding the action in the exponential to first order in $\sigma_i$ and using $\int \text{d}z_i w_i(z_i) \exp(\sigma_i \lambda_i^p z_i) = 1 + C^{\xi_i}(\sigma_i \lambda_i^p)$ (Eq.~\ref{eq:formerB8}) gives the Hamilton-Jacobi equation:
\begin{alignat}{1}
    - \frac{\partial S}{\partial t} = H
\end{alignat}
with the Hamiltonian:
\begin{equation}{}
    H = \sum_i \Big[v_i \lambda^\phi_i  - \alpha_i p_i \lambda^p_i + C^{\xi_i}(\sigma_i \lambda^p_i)
    + \big(- \gamma_i v_i + P_i + p_i + \sum_j K_{ij} \sin(\phi_j - \phi_i)\big) \cdot \frac{\lambda^v_i}{M_i} \Big].
\end{equation}
Note that in contrast to \cite{hindes_network_2019}, our Hamiltonian contains the cumulant generating function $C^{\xi_i}$ of the fluctuations $\xi_i$. The cumulants are related to the power increment fluctuations (generated by the Fourier-method) according to Eq.~\ref{eq: p_from_xi} and quantify non-Gaussian effects: for Gaussian noise all cumulants $\Gamma_n^{\xi_i}$ with $n > 2$ are zero.
The expansion of the cumulant generating function $ C^{\xi_i}(\sigma_i \lambda^p_i) = \sigma_i \Gamma_1^{\xi_i} \cdot \frac{\lambda_i^p}{1!} + \sigma^2_i \Gamma_2^{\xi_i} \cdot \frac{(\lambda_i^p)^2}{2!} + \dots $ can be alternatively obtained if the Kramers-Moyal expansion \cite{kampen_stochastic_1992}
 is applied to the differential Chapman-Kolmogorov equation (i.e. the Taylor expansion of $\pr(\bm{\phi}, \bm{v}, \bm{p} - \bm{\mathbbm{1}}_i \sigma_i z_i, t)$ in $\sigma_i z_i$).

The stationary solution $S_{stat}(\bm{\phi}, \bm{v}, \bm{p})$ to the Hamilton-Jacobi equation can be equivalently represented in terms of Hamilton's equations:
\begin{alignat}{3}
    S_{stat}(\bm{\phi}, \bm{v}, \bm{p}) = \int_{t_0}^{t_1} \dd t \sum_i \big[\lambda^\phi_i \frac{\dd \phi_i}{\dd t} + \lambda^v_i \frac{\dd v_i}{\dd t} + \lambda^p_i \frac{\dd p_i}{\dd t}] \,,
\end{alignat}
with $t_1 \gg t_0$ and where $\bm{\phi}(t)$, $\bm{v}(t)$, $\bm{p}(t)$, $\bm{\lambda}^{\bm{\phi}}(t)$, $\bm{\lambda}^{\bm{v}}(t)$ and $\bm{\lambda}^{\bm{p}}(t)$ are solutions of Hamilton's equations:
\begin{alignat}{4}
    \frac{\dd \phi_i}{\dd t} &=& \frac{\partial H}{\partial \lambda^\phi_i} &= v_i \,,\nonumber\\
    \frac{\dd v_i}{\dd t} &=&  \frac{\partial H}{\partial \lambda^v_i} &= \Big(- \gamma_i v_i +  P_i + p_i + \sum_j K_{ij} \sin(\phi_j - \phi_i)\Big)/M_i \,, \nonumber \\
    \frac{\dd p_i}{\dd t} &=&  \frac{\partial H}{\partial \lambda^p_i} &= - \alpha p_i(t) + \sigma_i C^{' \xi_i}(\sigma_i \lambda_i^p) \,, \nonumber\\
    \frac{\dd \lambda^\phi_i}{\dd t} &=& - \frac{\partial H}{\partial \phi_i} &= -\sum_j \big( K_{ij} \cos(\phi_j - \phi_i) \cdot [\lambda^v_j/M_j - \lambda^v_i/M_i] \big) \,,
    \nonumber\\
    \frac{\dd \lambda^v_i}{\dd t} &= & - \frac{\partial H}{\partial v_i} &= - \lambda^{\phi}_i + \gamma_i \lambda^v_i/M_i \,,
    \nonumber\\
    \frac{\dd \lambda^p_i}{\dd t} &=& - \frac{\partial H}{\partial p_i} &= \alpha \lambda^p_i - \lambda^v_i/M_i \,, \label{eq: hamiltons_eqs}
\end{alignat}
with all the parameters as in the swing equations.
Here $C^{' \xi_i}$ denotes the derivative of the cumulant generating function of $\xi_i$. While in our Hamilton's equations the higher order cumulants naturally occur, it is higher order moments in the approach of \cite{hindes_network_2019}. Furthermore note that Hamilton's equations for $\bm{\phi}$, $\bm{v}$ and $\bm{p}$ are the same as the swing equations with the noise terms $\sigma_i \xi_i$ replaced by $C^{' \xi_i}(\sigma_i \lambda_i^p)$, with the momenta $\lambda_i^p(t)$ determined by the remaining equations.
{} \\
Hamilton's equations are subject to the boundary conditions that the system is at the stable fixed point $\bm{\phi}^{\ast}$ (and other variables equal to $0$) at time $t_0$ and at $(\bm{\phi}, \bm{v}, \bm{p})$ (and other variables equal to $0$) at time $t_1$. The mean time of escape via the saddle $\bm{\phi}^s$ is then estimated as \cite{hindes_network_2019,dykman1,assaf_wkb_2017}
\begin{alignat}{1}
    \log \langle T \rangle = \text{constant} + S(\bm{\phi}^s, 0, 0) \,, \label{eq: T_from_S}
\end{alignat}
with the constant independent of the strength of the fluctuations (though it does depend on other parameters).
{} \\
Let us consider the special case of Gaussian noise. For Gaussian noise, $C_i'(\lambda_i^p) = \sigma^2_i \lambda_i^p$, moreover, $S \propto 1/(\bm{\sigma}\cdot \bm{\sigma})$ \cite{touchette_large_2009,freidlin_random_2012,dembo_large_2010}. The proportionality of the action to the inverse of the noise strength can still be seen from Hamilton's equations by rescaling the noise strength by some constant $c$, i.e. $\sigma^2_i \rightarrow {\sigma'}_i^2 = c \cdot \sigma^2_i$ at all nodes. For Gaussian noise, Hamilton's equations are linear in the momenta. If $\lambda^\phi_i$, $\lambda^v_i$, $\lambda^p_i$ solve Hamilton's equations for noise strengths $\sigma^2_i$, then ${\lambda'}^\phi_i = \lambda^\phi_i/c$, ${\lambda'}^v_i = \lambda^v_i/c$, ${\lambda'}^p_i = \lambda^p_i/c$ (and $\bm{\phi}$, $\bm{v}$, $\bm{p}$ unchanged) solve Hamilton's equations for noise strengths ${\sigma'}^2_i$, thereby changing the action:
\begin{alignat}{1}
    S \rightarrow S'  &= \int_{t_0}^{t_1} \dd t \sum_i \big[{\lambda'}^\phi_i \frac{\dd \phi_i}{\dd t} + {\lambda'}^v_i \frac{\dd v_i}{\dd t} + {\lambda'}^p_i \frac{\dd p_i}{\dd t} \big] \\
    &= \frac{1}{c}\int_{t_0}^{t_1} \dd t \sum_i \big[\lambda^\phi_i \frac{\dd \phi_i}{\dd t} + \lambda^v_i \frac{\dd v_i}{\dd t} + \lambda^p_i \frac{\dd p_i}{\dd t} \big] \\
    &= S/c \,.
\end{alignat}
Thus, the action is proportional to the inverse of the noise strength.

\subsubsection{Hamilton's equations in reduced phase spaces}\label{subsec45}
When the WKB approach is applied to the subgraph approximation as introduced for the swing equations in Sec.~\ref{subsec42}, Hamilton's equations of motion are given by Eqs.~\ref{eq: hamiltons_eqs} with $i\in\{1,2\}$, labeling the two disjoint areas in which the system splits upon an overload. The parameters $P_i,\gamma_i,M_i, \sigma_i$ are given again by Eqs.~\ref{eq: xi_synchronized} (with $P_1 = - P_2 = P$ and $K_{12} = K$). The terms representing the presence of noise are given by $\sum_{i\in A_k} C^{\prime \xi_i} (\sigma_i\lambda_k^p) = C^{\prime \xi_k}(\sigma_k \lambda_k^p)$ with $k\in\{1,2\}$ for the two areas, where we use again that the sum of cumulants is the cumulant of sums (Eq.~\ref{eq: cumulants_ssa}). (The same kind of commutativity does not hold for the moments.) With the three variables $(\bm{\phi}, \bm{v}, \bm{p})$, each being two-component vectors, and the associated auxiliary momenta, the phase space becomes 12-dimensional. Numerical integration of Hamilton's equations of motion by means of the iterative action minimization method (IAMM) \cite{lindley_iterative_2013} turns out to be challenging and time-consuming due to the existence of multiple time-scales in the evolution of the momenta. We describe this method in Sec.~\ref{sec: IAMM}.
{} \\
A further reduction of phase space, disregarding the dynamics of the phase angles and their time derivatives altogether, is suggested by the fact that operators of the power grid attribute system splits to lines being overloaded, rather than due to other transient behaviour of the phase angles \cite{machowski_power_2008,ENTSOEreport}. Of course, in isolation this argument does not justify to drop the phase dynamics, but we will further justify this approximation in Sec.~\ref{sec: overload}. Besides $P = (P_1 - P_2)/2$, the difference in power input in the two regions involved in the system split is quantified by:
\begin{alignat}{1}
p(t) \equiv (p_1(t) - p_2(t))/2 \,,
\end{alignat}
with time-evolution:
\begin{alignat}{1}
    \dot{p} &= - \alpha p + \sigma \xi \,, \label{eq: p}\\
\sigma \xi &\equiv (\sigma_1 \xi_1 - \sigma_2 \xi_2)/2 \label{eq: xi_hitting} \\
    \sigma^2 &\equiv \frac{\sigma_1^2 + \sigma_2^2}{4} \,, \label{eq: sig_hitting}
\end{alignat}
with areas $1$ and $2$ as in the subgraph approximation of Sec.~\ref{subsec42}. An overload is hit when $P + p(t) = K$. The associated reduction in Hamilton's equations is obtained by dropping the dynamics of $\bm{\phi}, \bm{v}, \bm{\lambda^\phi},\bm{\lambda^v}$:
\begin{alignat}{1}
    \frac{\mathrm{d} p}{\mathrm{d}t} &= - \alpha p + \sigma C^{'\xi}(\sigma \lambda^p) \,, \label{eq: p_hamil_reduced}\\
    \frac{\mathrm{d}\lambda^p}{\mathrm{d}t} &= \alpha \lambda^p \,, \label{eq: l_hamil_reduced}\\
    S &= \int_{-\infty}^{{t_o}} \mathrm{dt}  \lambda^p \dot{p} \,, \label{eq: S_reduced}
\end{alignat}
where $C^{'\xi}(\sigma\lambda)$ is the derivative of the cumulant generating function of $\xi$, and with boundary conditions $p({t_o}) = K -P$, and $p(-\infty) = \lambda^p(-\infty)  = 0$, where $t_o$ denotes the time at which the overload occurs. The system of equations is invariant under shifting of the time axis, and hence the action $S$ will be independent of ${t_o}$. Hamilton's equations are then solved for $t < {t_o}$ (for $t > {t_o}$ the variables are assumed to follow their deterministic trajectories), by:
\begin{alignat}{1}
    p(t) &= \int_{-\infty}^t \mathrm{d}t' \exp(- \alpha (t - t')) \sigma C^{'\xi}(\sigma \lambda^p(t')) \,,\\
    \lambda^p (t) &=   \lambda^p({t_o}) \exp(\alpha (t - {t_o}))\,,
\end{alignat}
where $\lambda^p({t_o})$ is found from Eq.~\ref{eq: p_hamil_reduced} by expanding the cumulant generating function:
\begin{alignat}{1}
    p({t}) &= \int_{-\infty}^{{t}} \mathrm{d}t' \exp\big(-\alpha (t - t')\big) \sum_{n = 2}^{\infty} \sigma^n\Gamma^{\xi}_{n} \frac{\lambda^p(t')^{n-1}}{(n-1)!} \nonumber\\
    &= \sum_{n = 2}^{\infty} \sigma^n\Gamma^{\xi}_{n} \int_{-\infty}^{{t}} \mathrm{d}t' \exp\big(\alpha (t' - t)\big) \frac{ \lambda^p({t})^{n-1} \exp\big(\alpha \cdot (n-1) \cdot (t' - t)\big)}{(n-1)!} \nonumber \\
    &= \sum_{n = 2}^{\infty} \sigma^n \Gamma^{\xi}_{n} \frac{\lambda^p({t})^{n-1}}{\alpha n!} \, .\label{eq: p_from_lambda} 
\end{alignat}
Setting $p({t_o}) = K - P$, this equation can be solved (numerically) for $\lambda^p({t_o})$.
Integrating the action of Eq.~\ref{eq: S_reduced} term by term (where it is convenient to substitute $\lambda^p$ as integration variable and use Eqs.~\ref{eq: p_hamil_reduced}-\ref{eq: l_hamil_reduced} and \ref{eq: p_from_lambda}), for Gaussian noise the result is:
\begin{alignat}{1}
S &=  \alpha \Big[\frac{K - P}{\sigma}\Big]^2 \,,\label{eq: gaussian_action}
\end{alignat}
while for non-Gaussian noise the action (containing Eq.~\ref{eq: gaussian_action} as a special case) gives
\begin{alignat}{1}
    S = \sum_{n = 2}^\infty \frac{[\lambda^p({t_o})]^n}{\alpha n!} \frac{n-1}{n} \sigma^n \Gamma^{\xi}_{n} \label{eq: non-gaussian_action}. 
\end{alignat}
Thus, for the two-dimensional phase space $(p, \lambda^p)$ we have an analytical expression for the action  up to the fact that Eq.~\ref{eq: p_from_lambda} is solved numerically.

\subsubsection{Expansion of the action in the distance to overload}\label{secadditional}
The action of Eq.~\ref{eq: non-gaussian_action} can be expanded in $\frac{K-P}{\sigma}$ by inserting $\lambda^p(t_o) = c_1 \Big[\frac{K - P}{\sigma}\Big] + c_2 \Big[\frac{K-P}{\sigma}\Big]^2 + \dots$ (where constants $c_1, c_2, \dots$ are to be determined from Eq.~\ref{eq: p_from_lambda}). The lowest two orders are:
\begin{alignat}{1}
    S = \frac{1}{\alpha}\Big[\Big(\alpha \frac{K - P}{\sigma}\Big)^2 - \frac{4}{9} \Gamma^{\xi}_{3} \Big(\alpha \frac{K - P}{\sigma}\Big)^3 + \text{h.o.} \Big]\,, \label{eq: S_expansion}
\end{alignat}
with in general the lowest order contribution of $\Gamma^{\xi}_n$ being at order $\frac{1}{\alpha}\Big(\alpha \frac{K - P}{\sigma}\Big)^{n}$, such that when desynchronization events become very rare (large $\frac{K - P}{\sigma}$) the average desynchronization time $\langle T \rangle$ becomes increasingly sensitive to higher order cumulants (which require increasingly large amounts of data to accurately estimate them \cite{chan_higher-order_2020}), corresponding to non-Gaussian effects. This explains why the authors of \cite{katrin}, studying the regime where desynchronization are common, found that for their systems non-Gaussianity had only a small effect on desynchronization times. A large correlation time (small $\alpha$) suppresses the higher order terms, so that $\alpha \frac{K - P}{\sigma}$ is a good expansion parameter. Eq.~\ref{eq: S_expansion} can be used to determine how the contribution of the skewness $\Gamma^{\xi}_3$ affects the desynchronization time. In terms of the stochastic fluctuations $\{p_i(t)\}$ entering in the full network, $\Gamma^{\xi}_3$ is found from Eqs.~\ref{eq: xi_synchronized}, \ref{eq: xi_hitting} and \ref{eq: alphatau_approx} as:
\begin{alignat}{1}
    \Gamma^{\xi}_3 = \frac{1}{2^3 \sigma^3 \tau}\Big[\sum_{i \in A_1}\Gamma_3^{p_i(t + \tau) - p_i(t)} - \sum_{j \in A_2} \Gamma_3^{p_j(t + \tau) - p_j(t)} \Big] \cdot [1 +  O(\alpha \tau)]\,, \label{eq: third_cumulant}
\end{alignat}
where $A_1$ is the area with positive power production $\sum_{i \in A_1} P_i$ and $A_2$ is the area with negative power production $\sum_{j \in A_2} P_j = -\sum_{i \in A_1} P_i$. Eqs.~\ref{eq: S_expansion}-\ref{eq: third_cumulant} therefore show that whether a third cumulant (skewness) in the noise input increases or decreases the time to desynchronization depends on where in the network it enters: if it enters in the area with positive power input (here area $A_1$) a positive (negative) skewness decreases (increases) the average time to desynchronization, while if it enters in the area with negative power input (here area $A_2$) a positive (negative) skewness increases (decreases) the average time to desynchronization.

The authors of \cite{hindes_network_2019}, assuming the parameters are spatially homogeneous, made an expansion similar to Eq.~\ref{eq: S_expansion}. Their expansion is formally in $\frac{K - P}{P}$, parameterizing the distance to the bifurcation point, and the main difference in the assumptions on their network parameters is that they set the fluctuation timescale to $1/\alpha = 1 \,\text{s}$ (while here we set it, in accordance with \cite{tyloo_finite-time_2022}, two orders of magnitude larger) \footnote{Formally, for spatially homogeneous systems their expansion is valid in the limit where $\frac{K - P}{P}$ is much smaller than all other relevant quantities, but for the realistic values we consider here we have $\frac{M \alpha^2}{|P|} \ll \frac{K - P}{P} $ and $\frac{\gamma \alpha}{|P|} \ll \frac{K - P}{P}$, invalidating their expansion in our case.}.
Thus, our approximations may not be accurate for the cases studied in  \cite{hindes_network_2019} and vice-versa.
Nevertheless, the expressions of \cite{hindes_network_2019} (for spatially homogeneous parameters) can be directly compared to the expansion of Eq.~\ref{eq: S_expansion} if we consider both as an expansion in $K$ around $K = P$, that is, an expansion in the distance to the bifurcation point, as summarized in Table~\ref{tab: comparison}. Qualitatively, their predicted influence of the third cumulant is the same as we found above.

However, there are also significant differences in the two expansions; in particular, the lowest order contribution to the action found in \cite{hindes_network_2019} is at order $(K - P)^{3/2}$, and is proportional to $\gamma$ and $\alpha^2$. In contrast, the lowest order contribution to the action of Eq.~\ref{eq: non-gaussian_action} is at order $(K - P)^2$, is proportional to $\alpha$ and completely independent of $\gamma$. For Eq.~\ref{eq: non-gaussian_action} the order $(K- P)^2$ is exact for the Gaussian case, while in the results of \cite{hindes_network_2019} all orders are relevant also for the Gaussian case. Lastly, in Eq.~\ref{eq: non-gaussian_action} the lowest order at which the $n$-th cumulant enters the action is $O\big((K - P)^{n} \big)$, while for \cite{hindes_network_2019} this is at $O\big((K - P)^{n - 1/2} \big)$. These differences illustrate the sensitivity to different analytical approaches and the importance of the correlation time $1/\alpha$ of the stochastic fluctuations (which a simple white noise process for $p$, corresponding to $1/\alpha \rightarrow 0$, would not include). Further work should therefore extend this investigation of the interplay between non-Gaussianity and time-correlations also to more realistic power spectra \cite{katrin}.

\begin{table}
\centering
    \begin{tabular}{ |l|c|c| }
     \hline
     \begin{tabular}{c} \\ \end{tabular} &Eq.~\ref{eq: non-gaussian_action} &  \cite{hindes_network_2019} \\
      \begin{tabular}{c}  Lowest order contribution \\ \end{tabular} & $\frac{\alpha }{\sigma^2} (K - P)^2$ &$\propto \frac{\gamma \alpha^2}{\sqrt{P} \sigma^2} (K- P)^{3/2}$\\
      \begin{tabular}{c} Lowest order at which \\ $\Gamma^{\xi}_n$ contributes \end{tabular}& $O\Big(\Big[\alpha \frac{K - P}{\sigma}\Big]^{n}\Big)$ & $O\Big(\Big[\frac{K - P}{P}\Big]^{n - 1/2} \Big)$\\
      \begin{tabular}{c} Lowest order impact \\ of positive skewness \end{tabular} & \begin{tabular}{c} Increase of $\langle T \rangle$ in area\\ with positive generation \\ and vice-versa \end{tabular} & \begin{tabular}{c} Increase of $\langle T \rangle$ in the area\\ with positive generation \\ and vice-versa \end{tabular}\\
     \hline
    \end{tabular}
    \caption{Comparing the action of Eq.~\ref{eq: non-gaussian_action} to the expressions of \cite{hindes_network_2019}.}
    \label{tab: comparison}
\end{table}
{}

\section{Methods}\label{sec5}
\subsection{Sampling of the swing equations with the Fourier-method}
The pedestrian way to calculate the average desynchronization time is to numerically integrate the swing equations Eqs.~\ref{eq: swing_dynamics} with fluctuation dynamics Eq.~\ref{eq: p_dynamics}, choosing $Z_{\Delta t,i}$ of Eq.~\ref{eq:17} with probability $\text{pr}(Z_{\Delta t,i})$, determined by the Fourier-method, that is, calculating $\text{pr}(Z_{\Delta t,i})$ as inverse Fourier transformation of Eq.~\ref{eq:24}. For the subgraph approximation of the swing equations, the sampling refers to the same method, but applied to Eqs.~\ref{eq: reduced_eq_1}-\ref{eq: xi_synchronized}.

Here, we calculate the average desynchronization time, for given parameters, by calculating an average over $6000$ samples. The CPU time necessary for this computation is roughly proportional to the average desynchronization time. If the average desynchronization time is $\sim 100$ minutes, calculating the average desynchronization time for the given set of parameters (i.e. for each of the points shown in the figures) takes $\approx 200$ hours of CPU time on an Intel E5-2640 processor.  Doing the same computation for an average desynchronization time of $\sim 1$ year would then cost $\approx 10^6$ hours of CPU time, per set of parameters, which would clearly be infeasible. To capture really rare blackouts, one should resort to the WKB-approach. Nevertheless, in a range where both methods apply and are feasible, the WKB approach should be supplemented by sampling to validate the results of the various approximations within the WKB approach, to calculate the constant in $\log \langle T \rangle$ (which is not predicted by WKB), to find how rare events should be for the WKB approach to provide an accurate approximation, and, if of interest, to capture also the regime where desynchronization events are not rare, where WKB does not apply.

\subsection{The iterative action minimization method} \label{sec: IAMM}
 To solve Hamilton's equations of motion for the synchronized subgraph approximation, we have to deal with the 12-dimensional phase space. Here we use the iterative action minimization method \cite{lindley_iterative_2013} to numerically find the solutions. The method requires a discretization of Hamilton's equations. We take a time interval $[T_0, T_1]$, and discretize time into $n = 200$ points along a grid $\{t_i: i = -n/2 + 1, -n/2 + 2, \dots, n/2 \}$ with:
\begin{equation}
    t_i = T_0 + (T_1 - T_0) \frac{\sum_{j =1}^i [\cosh(a \cdot j )^2/(\cosh(a \cdot j )^2 + b -1)]}{\sum_{j = 1}^n [\cosh(a \cdot j )^2/(\cosh(a \cdot j )^2 + b -1)]} \,.
\end{equation}
For our concrete case we used  $T_1 = - T_0 = 1075$, $a = 0.055$, $b = 10$.\\
At grid points $i = 5, \dots, n - 4$, we approximate the derivatives that occur in Hamilton's equations by $8$-th order central differences \cite{singh_finite_2009}:
\begin{align}
    \frac{\mathrm{d} f(t)}{\mathrm{d}t} \biggr\rvert_{t_i} &\approx  \sum_{j = i -4}^{i + 4} c_{ij} f(t_j) \,,\\
    c_{ij} &\equiv \begin{cases}
                \sum_{l \in (i-4, \dots, i + 4) \setminus i} \frac{1}{t_i - t_l} &\text{if} \quad j = i\ \,, \vspace{0.5cm} \\
                \frac{\prod_{k \in (i-4, \dots, i + 4) \setminus i,j} (t_i - t_k)}{\prod_{k \in (i-4, \dots, i + 4) \setminus j} (t_j - t_k)} & \text{else} \,.
            \end{cases}
\end{align}
For $i = 1,2,3, 4, n-3, n-2, n-1, n$ we replace Hamilton's equations by the constraints that all the variables equal the boundary conditions, e.g., $\bm{\phi}(t_1) = \bm{\phi^\ast}$ and $\bm{\phi}(t_n) = \bm{\phi^s}$ for the phase angles. Here $\bm{\phi^\ast}$ and $\bm{\phi^s}$ are only determined up to a global phase shift, which we set according to Eqs.~\ref{eq: phi_ast}-\ref{eq: phi_s}. Together, this discretizes Hamilton's equations into a set of $2 \cdot 6 \cdot n$ equations, which can be solved with an off-the-shelf solver. We used the Levenberg-Marquardt algorithm implemented in Scipy \cite{more_levenberg-marquardt_1978, virtanen_scipy_2020}. The method requires an initial guess; we find that for the parameters of the Brazilian network of Sec.~\ref{subsec22} and $1/\alpha = 100 \text{s}$ the algorithm does not converge without a very accurate guess.

\begin{figure}
    \centering
    \subfloat[]{\includegraphics[width = 0.5 \textwidth]{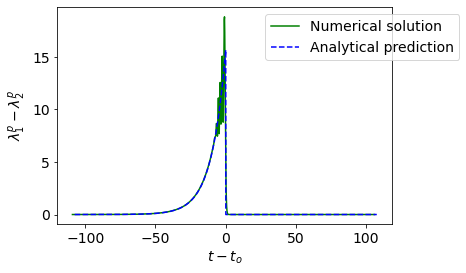}}
    \subfloat[]{\includegraphics[width = 0.5 \textwidth]{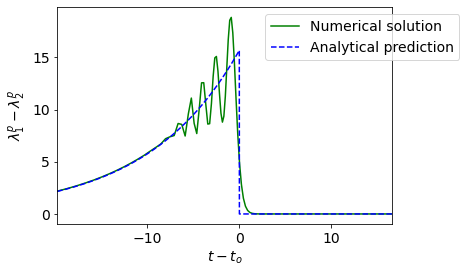}}
    \caption{Momentum $\lambda^p$ found numerically by the IAMM (zoom-in in b), compared to that predicted by the overload approximation. The numerical solution shows oscillations near the overload, presumably due to the rapid drop to zero after the overload. Here $t_o$ denotes the time at which the phase difference reaches $\frac{\pi}{2}$ (Hamilton's equations are invariant under a time translation and hence $t_o$ is not fixed by the equations).}
    \label{fig: example_momentum}
\end{figure}

To overcome this, we first artificially multiply all $\gamma_i$ values by a factor thousand. As a guess we then use the near-bifurcation solution given in \cite{hindes_network_2019} (this guess needs spatially homogeneous parameters, for which we insert the average values of the actual parameters), and use the solution as a new guess for slightly lower values of $\gamma$, until the actual values are reached. As a further slight improvement to the accuracy of the solution, we afterwards increase the discretization to $n = 800$, using the solution for $n = 200$ (linearly interpolated to the extra gridpoints) as an initial guess.\\
Decreasing the damping $\gamma$ too quickly leads to oscillations in the numerical solution for the momenta (and eventually for all the variables). If the values are lowered too fast, oscillations in the solution will increase and prevent convergence. An example is shown in Fig.~\ref{fig: example_momentum}. The numerical solution shows that, as a function of $t$, momentum builds up over time, and then -after the overload is hit- very quickly drops to $0$. The difference in these two timescales creates problems with the discretization, as we would need to discretize time in a very large number of steps to capture both scales accurately. This leads to oscillations in the numerical solution of the momentum near the overload. The damping should be decreased in small ($\approx 50$) steps, and Hamilton's equations solved numerically take up to several hours for low damping each time. This makes this approach infeasible for exploring a large space of parameters. Nevertheless, as we will now see, the optimal path that is obtained with this approach is still useful as a justification for the overload approximation of Sec.~\ref{subsec45}.
\subsection{Justification for the overload approximation} \label{sec: overload}
The overload approximation amounts to a restriction of the dynamics to that of the power fluctuations. Combined with the WKB-method this led to Hamilton's equations of motion in two-dimensional phase space according to Eqs.~\ref{eq: p_hamil_reduced}-\ref{eq: l_hamil_reduced}. Looking purely at the mathematical expressions, what would suggest this reduction? If we start with the subgraph approximation and
subtract Eq.~\ref{eq: reduced_eq_2} from Eq.~\ref{eq: reduced_eq_1}, we obtain:
\begin{alignat}{1}
(M_1 \dot{v}_1 - M_2 \dot{v}_2 )/2 + (\gamma_1 v_1 - \gamma_2 v_2)/2 &=  P + p(t) + K \sin(\phi_2 - \phi_1) \,, \label{eq: reduced_eq_5}
\end{alignat}
with $p(t) \equiv (p_1(t) - p_2(t))/2$. The scaling of Eq.~\ref{eq:10} shows that $\dot{v} \sim \alpha^2$ and $v \sim \alpha$; more precisely, since $M_i$ and $\gamma_i$ have units of $\text{time}^2$ and $\text{time}$, respectively, only their values relative to the timescale $1/\alpha$ matter.  The fact that here $\frac{M_i\alpha^2}{P} \sim 10^{-6}$ and $\frac{\gamma_i \alpha}{P} \sim 10^{-3}$ (Sec.~\ref{subsec22}) motivates the assumption that the left-hand side of Eq.~\ref{eq: reduced_eq_5} is negligible along the optimal path in the WKB-method. This amounts to a time scale separation between the phase dynamics and the correlation time between power increment fluctuations. In the WKB-approach this means that the dynamics gets independent of inertia and damping. The result for inertia is in agreement with \cite{tyloo_finite-time_2022} and \cite{tyloojacquod}, according to which inertia has little impact on the voltage angle fluctuations for long correlation times.

\begin{figure}
    \centering
    \subfloat[]{\includegraphics[width = 0.5 \textwidth]{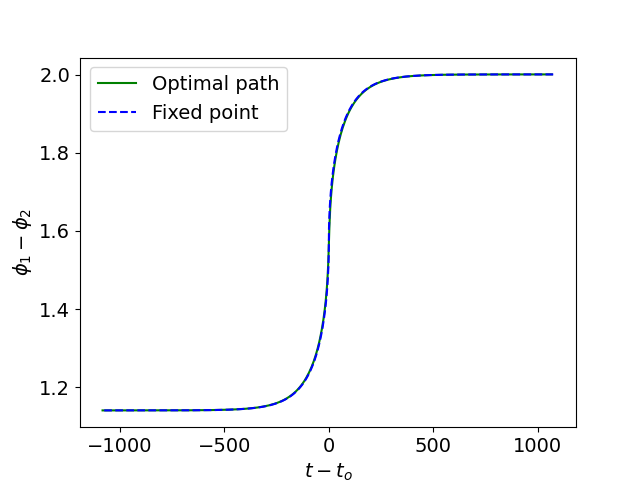}}
    \caption{The phase difference between the two areas involved in the system split (here reduced to two oscillators), along the optimal path to desynchronization (obtained by a numerical solution of Hamilton's equations, Sec.~\ref{subsec45}) for the Brazilian power network (Sec.~\ref{subsec22}) with Gaussian noise input and distance to overload $\frac{K - P}{K} = 9.1 \%$ (the path is independent of $\sigma$), as compared to the fixed point of the swing equations with power input $P + p(t)$ (Eq.~\ref{eq: phi_path}). Here $t_o$ denotes the time at which the phase difference reaches $\frac{\pi}{2}$ (Hamilton's equations are invariant under a time translation and hence $t_o$ is not fixed by the equations).}
    \label{fig: WKB}
\end{figure}

The phase difference $\phi_2 - \phi_1$ then simply 
follows an equation that determines the fixed point of the swing equations, satisfying $p(t) + P + K \sin(\phi_2 - \phi_1) = 0$. For $P + p(t) < K$ (i.e. no overload) there are the two known fixed points (as in Eqs.~\ref{eq: phi_ast}-\ref{eq: phi_s}), the stable fixed point and the saddle-node. The stable fixed point itself agrees with the solution of the static power flow equations, here not further considered. \footnote{The fixed point solution of Eq.1 is obtained for the active power and voltage phases. It corresponds to the solution of the steady load flow if the reactive power and the dynamics of the voltage magnitude are neglected. On the engineering side, much effort was invested in determining a probabilistic load flow \cite{chenchen}. In contrast to these approaches, we are not interested in the full solution to the steady load flow under diverse sources of fluctuations, but only in the fluctuating power flowing between the two areas connected by the overloaded lines: In the WKB-approach, the probability for a desynchronization event is then easily calculated with much less effort than required in the probabilistic load flow approaches.} For $P + p(t) \rightarrow K$ the fixed points merge into one. If the line gets overloaded at some $t = t_o$, the system resides at the stable fixed point for all $t < t_o$ and resides at the saddle fixed point for $t > t_o$ (with the two fixed points merged at $t = t_o$), then the phase difference follows the trajectory:
\begin{alignat}{1}
\phi_1(t) - \phi_2(t) &=
 \begin{cases}
    \arcsin(\frac{P + p(t)}{K})  \quad &\text{for} \quad t < {t_o}\, \\
    \pi/{2} \quad &\text{for} \quad t = {t_o}\, \\
    \pi - \arcsin(\frac{P + p(t)}{K})  \quad &\text{for} \quad t > {t_o}\,.
\end{cases} \label{eq: phi_path}
\end{alignat}
Fig.~\ref{fig: WKB} shows that the optimal path computed numerically from Hamilton's equations for the synchronized subgraph approximation (Sec.~\ref{sec: IAMM}) is indistinguishable from the path predicted by Eq.~\ref{eq: phi_path}, such that indeed neglecting the left-hand side of Eq.~\ref{eq: reduced_eq_5} is justified at the level of the action.

The authors of \cite{tyloo_finite-time_2022, zhang_fluctuation-induced_2019} show that, for typical fluctuations, dropping the phase dynamics is always valid in the regime of long correlation time. In contrast, we argue that for rare events this time-scale separation, here leading to the overload approximation, should only be seen as an approximation at the level of the action (Eqs.~\ref{eq: T_from_S}, \ref{eq: non-gaussian_action}) and the optimal path (Fig.~\ref{fig: WKB}) rather than of the full desynchronization time (the full desynchronization time would include the prefactor of the exponential).\footnote{This is unlike the synchronized subgraph approximation (whose partition of the grid is suggested by the values of the slowest mode of the system, the Fiedler mode), which approximates both small and large fluctuations. Both approximations turn out to also apply away from the bifurcation point.} This means it gives the order of magnitude of the average time to desynchronization only up to exponential accuracy, even for very large correlation time of the power input.

This can be seen as follows.
If we consider the dynamics of $p$ according to Eq.~\ref{eq: p} in isolation, for Gaussian noise $p$ follows an Ornstein-Uhlenbeck process \cite{ricciardi_first-passage-time_1988}. The results of \cite{ricciardi_first-passage-time_1988} then give that in the limit of large $(K-P)/\sigma$ the expected hitting time for $p \rightarrow K - P$ follows:
\begin{alignat}{1}
    \langle t_o \rangle = \frac{\sqrt{\pi}}{\alpha^{3/2}}\frac{\sigma}{(K -P)}\exp\Big(\alpha \Big[\frac{K - P}{\sigma}\Big]^2 \Big) \,, \label{eq: hitting}
\end{alignat}
where the exponent agrees with Eq.~\ref{eq: gaussian_action}. This suggests the approximation $\langle t_o \rangle \approx \langle T \rangle$ for the expected desynchronization time. However, in escape problems with Gaussian noise the prefactor (corresponding to the constant in Eq.~\ref{eq: T_from_S}) is known to depend sensitively on the dynamics near the boundary of the basin of attraction \cite{bouchet_generalisation_2016}. For the swing equations, the dynamics of $p$ alone cannot be expected to give a good approximation to the full system near the overload: For $p(t) \rightarrow (K - P)$,  Eq.~\ref{eq: phi_path} predicts $(v_1 - v_2) \rightarrow \infty$, such that close to overload neglecting the left-hand side of Eq.~\ref{eq: reduced_eq_5} is not justified. The prefactor in Eq.~\ref{eq: hitting} is valid for an isolated Ornstein-Uhlenbeck process, but as a prediction for the desynchronization time of the swing equations it should be taken with care; in Sec.~\ref{sec: results} we show that this analytical formula for the overload hitting time does not give an accurate approximation for the prefactor of the desynchronization time $\langle T \rangle$, when compared to the values of $\langle T \rangle$ calculated with the sampling of the full swing equations. Still, it is in the right ballpark. \\
\section{Results} \label{sec: results}
\subsection{Comparison of the Fourier method and the compound Poisson process for data implementation}
\begin{figure}
    \centering
    \subfloat[]{\includegraphics[width = 0.45 \textwidth]{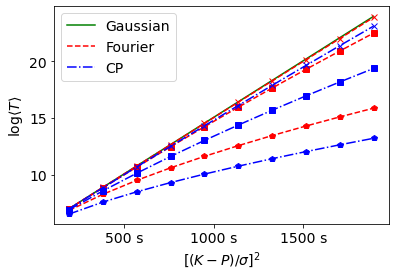}}
    \subfloat[]{\vspace{-1cm}\includegraphics[width = 0.55 \textwidth]{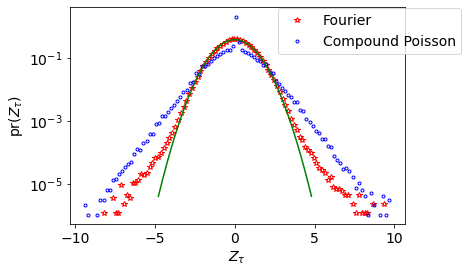}}
    \caption{(a) Impact of the data implementation on the average desynchronization time according to the Fourier (compound Poisson) method, red (blue) color, respectively, for three values of $1/\alpha$: from top to bottom, $1/\alpha = 10^2 \, \text{s}, 10^1 \, \text{s}, 10^0 \, \text{s}$ (crosses, squares, circles). The Gaussian case (green) is reached for ($1/\alpha\rightarrow \infty$). (b) Histograms of standardized increments $Z_\tau$ for $5 \cdot 10^6$ samples, $10^3$ bins (where the Fourier-method samples from the exact distribution $\text{pr}(Z_\tau = z)$), colors as in (a). For further explanations see the text.}
    \label{fig: fourier_vs_poisson}
\end{figure}

In the following we assume that the approximation used for the Fourier method according to Eq.~\ref{eq:fourier} gives an accurate approximation of the data.
For the construction of a compound Poisson process, the relative error on the cumulants of the increments induced by setting $\rho = 1/\tau$ and $\lambda(z) = \pr(Z_{\tau,i}  = z)$ is then quantified 
by $\Gamma_n^{Z^{\text{cp}}_\tau}/\Gamma_n^{Z_{\tau,i}} - 1 = \mu_n^{\lambda}/\Gamma_n^{Z_{\tau, i}} - 1 = \mu_n^{Z_{\tau, i}}/\Gamma_n^{Z_{\tau,i}} - 1 = \mu_n^{\text{data}}/\Gamma_n^{\text{data}} - 1$, where the equality signs follow from Eq.~\ref{eq: poiss_moments}, $\lambda(z) = \pr(Z_{\tau,i}  = z)$, and Eq.~\ref{eq: data_identification}, respectively, which can be evaluated from the data on the increment distribution. Just to recapitulate: $\Gamma_n^{Z^{\text{cp}}_\tau}$ refer to the n-th cumulants of the power increments of the compound Poisson process, $\Gamma_n^{Z_{\tau,i}}$ to the cumulants of the power increments over the interval $\tau$, identified with those of the data $\Gamma_n^{\text{data}}$, $\mu_n^{\lambda}$  to the n-th moments of $\lambda_i(z)=w_i(z)$, $\mu_n^{Z_{\tau, i}}$ and $\mu_n^{\text{data}}$ accordingly.  This means that we can take the difference between moments and cumulants of the data as indicating the error in the reproduction of histograms by a compound Poisson process, if we assume Eq.~\ref{eq: data_identification}  gives an accurate approximation of the data. The errors on the first few cumulant ratios are shown in Table~\ref{tab: data_poisson_error}.
\begin{table}[b]
    \caption{\label{tab: data_poisson_error}%
    The relative error $\Gamma_n^{Z^{\text{cp}}_\tau}/\Gamma_n^{Z_\tau} - 1$ between cumulants of the increments $Z_\tau^{\text{cp}}$ and $Z_\tau$, induced by setting $\rho = 1/\tau$ and $\lambda(z) = \pr(Z_\tau  = z)$ in the compound Poisson process. Shown for various distributions of data for power increments, calculated as $\Gamma_n^{Z^{\text{cp}}_\tau}/\Gamma_n^{Z_{\tau}} - 1 = \mu_n^{\text{data}}/\Gamma_n^{\text{data}} - 1 $ (see Sec.~\ref{subsec31}). In particular, while the approximation is relatively accurate for the wind power generation data, it fails for Gaussian distributions, and gives significant errors on higher-order cumulants (e.g. $\sim 30\%$ for $ n = 12$) for the solar power generation data.
    }
    \begin{tabular}{| l|c c c c c c |}
    \hline
    Distribution& $n = 2$ & $4$ & $6$ & $8$ &$10$ & $12$ \\ \hline
    Gaussian  (zero-mean) & $0.$ & $\infty$ & $\infty$ & $\infty$ & $\infty$ & $\infty$\\
    Wind & $0.000$ & $0.021$ & $0.021$ & $0.025$& $0.034$ & $0.053$ \\
    Solar & $0.000$ & $0.024$& $0.049$ & $0.085$& $0.152$ & $0.302$ \\
    \hline
    \end{tabular}
\end{table}
The impact of these errors in the cumulants on the average desynchronization time are shown in
Fig.~\ref{fig: fourier_vs_poisson}.
As the difference is only clearly seen for very large times, we determine it within the overload approximation, for which the WKB approximation becomes fully analytically tractable. The implementation is either according to Sec.~\ref{subsec32} (the Fourier-method), or by a compound Poisson process as in Sec.~\ref{subsec31} ('CP'). We have chosen the noise close to a Gaussian distribution, as there the difference between the Fourier and compound Poisson approaches is largest:
\begin{equation}
\text{pr}(Z_\tau = z) = 0.9 \cdot N[0,1](z) + 0.05 \cdot \sqrt{2} \text{exp}(-\sqrt{2}|z|)\,,
\end{equation}
where $N$ denotes the normal distribution, with the parameters given in square brackets.
In the overload approximation, $\log \langle T \rangle$ is  predicted only up to a constant, which for comparison we here set equal to $5.05$ (which matches the results of sampling for Gaussian noise and $1/\alpha = 100 \;\text{s}$) for all cases. Recall that within the overload approximation noise enters via the cumulants $\Gamma^\xi_n$ in Eq.~\ref{eq: non-gaussian_action}. In the Fourier-method $\Gamma^\xi_n$ is approximated according to Eq.~\ref{eq: z_from_xi_cumulants} with $\Delta t=\tau$ and $\Gamma_n^{Z_{\tau,i}}$ identified with the cumulants of the data, while in the CP-method the cumulants are approximated by Eq.~\ref{eq: poiss_moments}  with $\Delta t=\tau$ and $\Gamma_n^{Z_{\tau}^{cp}}$ approximated by the moments of the data.

For the correlation time $1/\alpha \rightarrow \infty$, the difference to the results for a Gaussian distribution disappears, while the difference gets larger for small $1/\alpha$. In Fig.~\ref{fig: fourier_vs_poisson} we choose from top to bottom, $1/\alpha = 10^2 \, \text{s}, 10^1 \, \text{s}, 10^0 \, \text{s}$ (crosses, squares, circles).
Note that for large desynchronization times the difference between the two ways of including the noise gets larger. For example, the Fourier-method predicts the average desynchronization time to be $3.55$ days for $\Big(\frac{K - P}{\sigma} \Big)^2 = 760 \, \text{s}$ and $17.6$ years for $\Big(\frac{K - P}{\sigma} \Big)^2 = 1515 \, \text{s}$, both for $1/\alpha = 100 \;\text{s}$ (where differences are smallest), while the compound Poisson approximation predicts an average desynchronization time of $3.10$ days and $10.6$ years, respectively. If we compare the  increment data $Z_\tau$, artificially generated according to the Fourier-method or to the compound Poisson process, Fig.~\ref{fig: fourier_vs_poisson}(b) shows the different histograms of standardized increments $Z_\tau$ for $5 \cdot 10^6$ samples, $10^3$ bins (where the Fourier-method samples from the exact distribution $\text{pr}(Z_\tau = z)$ without the need of performing a Fourier transform).

\subsection{Comparison of the average desynchronization time for the different dimensional reductions}
\begin{figure}
    \centering
    \subfloat[]{\includegraphics[width = 0.5 \textwidth]{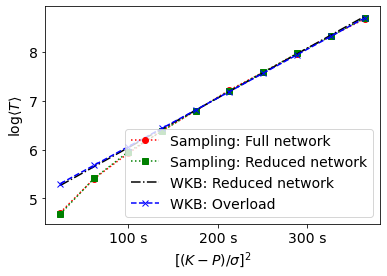}}
    \subfloat[]{\includegraphics[width = 0.5 \textwidth]{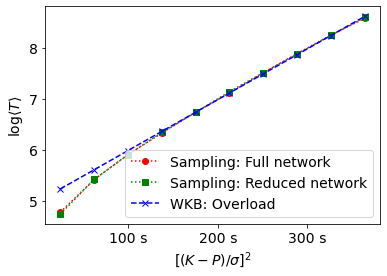}}
    \caption{Average desynchronization time $\langle T \rangle$ (in seconds) for Gaussian fluctuations on a model of the Brazilian power network (\ref{subsec22}) according to Fig.~\ref{fig: brazilian_network}(a), calculated either by direct simulation of the swing equations ($\delta t = 10^{-3}\,\text{s}$, $\langle T \rangle$ averaged over 6000 samples) for the full network and the reduced network, by the WKB method (Sec.~\ref{subsec45}) for the reduced network or by the overload prediction of Sec.~\ref{subsec45}. 
    (a) for constant distance to overload $(K - P)/K = 9.1 \%$ and varying $\sigma$, (b) for varying distances to overload $(K - P)/K = 5.9 - 22.8 \%$ and constant $\sigma$. For further explanations see the text.}
    \label{fig: escape_gauss}
\end{figure}

\begin{figure}
    \centering
    \subfloat[]{\includegraphics[width = 0.5 \textwidth]{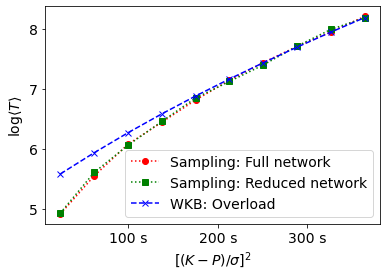}}
    \subfloat[]{\includegraphics[width = 0.5 \textwidth]{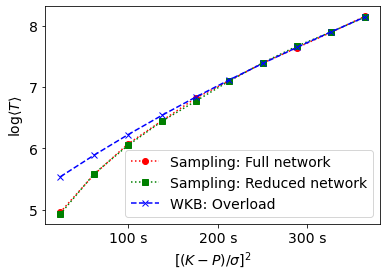}}
    \caption{Average desynchronization time $\langle T \rangle$ (in seconds) for fluctuations according to (rescaled) wind data on a model of the Brazilian power network (\ref{subsec22}) according to Fig.1(a), calculated either by direct simulation of the swing equations for the full network and reduced network or by the overload prediction of Sec.~\ref{subsec45} (Eqs.~\ref{eq: p_from_lambda}-\ref{eq: non-gaussian_action} truncated after $8$ cumulants). Otherwise the same as Fig.~\ref{fig: escape_gauss}.}
    \label{fig: escape_wind}
\end{figure}

In Figs.~\ref{fig: escape_gauss} (for Gaussian fluctuations)
and \ref{fig: escape_wind} (for wind data) we compare the average time to desynchronization between sampling of the full set of swing equations, sampling according to their subgraph approximation, the WKB-overload approximation, and, in Fig.~\ref{fig: escape_gauss}(a), the WKB-approximation in the subgraph approximation (12-dimensional phase space). Within the overload approximation of Sec.~\ref{subsec45}, the average desynchronization time only depends on $(K-P)/\sigma$ rather than on the individual quantities, as can be found by dividing Eq.~\ref{eq: p} by $\sigma$ or by considering the action Eq.~\ref{eq: gaussian_action}. Panels (a) and (b) in both figures differ in how desynchronization events become rare: (a) for constant distance to overload $(K - P)/K = 9.1 \%$ and varying $\sigma$ towards smaller values, (b) for increasing distances to overload $(K - P)/K = 5.9 - 22.8 \%$ and constant $\sigma$.
WKB predicts $\log \langle T \rangle$ only up to a constant (in general independent of $\sigma$, for the overload approximation also independent of $K - P$), which is here chosen to match the results by direct simulation. In all four panels we use the realization of the Brazilian grid according to Fig.~\ref{fig: brazilian_network}(a) and see very good agreement for $(K - P) \gg \sigma$. Sampling within the synchronized subgraph approximation is furthermore also an excellent approximation for events which are not rare, i.e. where the WKB approximation does not apply, and of the constant in $\log \langle T \rangle$ which is not predicted by the WKB approximation. Here $K$ is the conductance and $P$ the power flow through the most heavily loaded line, $K-P$ is the distance to overload, $\sigma$ is the noise strength (as in Eq.~\ref{eq: sig_hitting}), here chosen to be equal for both nodes at which stochastic fluctuations enter.

\begin{figure}
    \centering
    \subfloat[]{\includegraphics[width = 0.5 \textwidth]{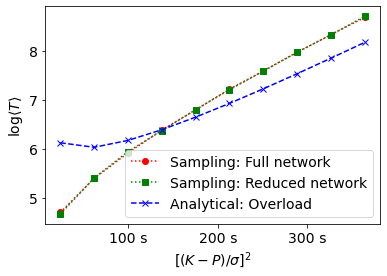}}
    \caption{Equivalent to Fig.~\ref{fig: escape_gauss},  with the average escape time $\langle T \rangle$ inaccurately (but not too badly) predicted by the full analytical formula (Eq.~\ref{eq: hitting}, including the prefactor) for the overload approximation for large $(K-P)/\sigma$ (Eq.~\ref{eq: hitting}) rather than by WKB. }
    \label{fig: hitting_time}
\end{figure}

In Fig.~\ref{fig: hitting_time} we compare the average time to desynchronization for the sampling of the full and reduced set of swing equations (with Gaussian noise) with the full analytical solution for the overload approximation according to Eq.~\ref{eq: hitting} (i.e. including the prefactor),
with the latter giving a somewhat inaccurate prediction. In contrast to the reduction by the synchronized subgraph approximation, the overload approximation is thus only valid at the level of the optimal path and the action, as already argued in Sec.~\ref{sec: overload}. \\
\begin{figure}
    \centering
    \subfloat[]{\includegraphics[width = 0.5 \textwidth]{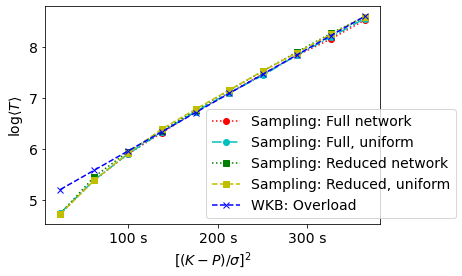}}
    \subfloat[]{\includegraphics[width = 0.5 \textwidth]{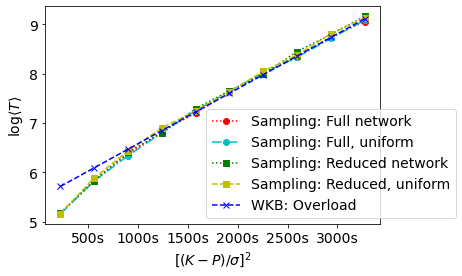}}
    \caption{Average desynchronization time $\langle T \rangle$ for Gaussian fluctuations on the model of Fig.~\ref{fig: brazilian_network}(b) and (c) ('uniform'), $1/\alpha= 100 \, \text{s}$, (b) $1/\alpha = 300\,\text{s}$. Otherwise the same as Fig.~\ref{fig: escape_gauss}(a).}
    \label{fig: switched_gauss}
\end{figure}
To consider a more interesting situation in which the overload is not along an edge to a leaf node,
we repeat in Figs.~\ref{fig: switched_gauss} and \ref{fig: switched_wind} the comparison between the various approximations for a realization of the Brazilian grid according to Fig.~\ref{fig: brazilian_network}(b) and (c), for which the network partitions into half along two lines. Additionally, in the network of Fig.~\ref{fig: brazilian_network}(c) we consider adding noise with the same distribution to all different nodes of the network, rather than noise entering only at two nodes. Figs.~\ref{fig: switched_gauss}(a)
and \ref{fig: switched_wind} show that the results obtained for the network of Fig.~\ref{fig: brazilian_network}(a) still hold with these adaptations. The analysis of \cite{tyloo_finite-time_2022} finds that appropriate values for $1/\alpha$ are $2$-$5$ minutes; to consider the higher end of this range, we consider $1/\alpha = 300\,\text{s}$ in Fig.~\ref{fig: switched_gauss}(b) and find that the same results hold as for $1/\alpha = 100\,\text{s}$ in Fig.~\ref{fig: switched_gauss}(a).
\begin{figure}
    \centering
    \subfloat[]{\includegraphics[width = 0.5 \textwidth]{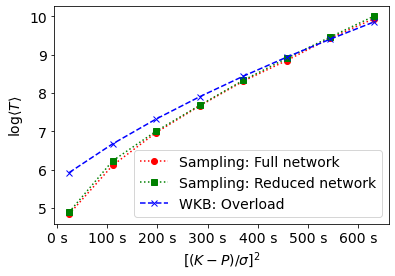}}
    \subfloat[]{\includegraphics[width = 0.5 \textwidth]{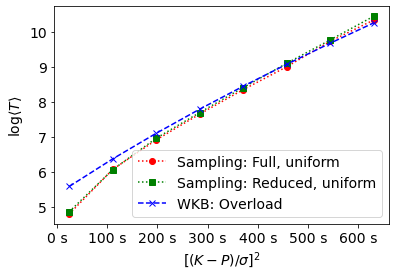}}
    \caption{Equivalent of Fig.~\ref{fig: switched_gauss} with fluctuations distributed according to the (rescaled) wind data, $\delta t = 10^{-4}\,\text{s}$, and each data point an average over $1800$ samples. (a) Fluctuations entering in two nodes (Fig.~\ref{fig: brazilian_network}(b)), (b) fluctuations entering at all nodes (Fig.~\ref{fig: brazilian_network}(c)).}
    \label{fig: switched_wind}
\end{figure}

\subsection{Skewness of the action: more or less desynchronization events for non-Gaussian data}
\begin{figure}
    \centering
    \includegraphics[width = 0.5 \textwidth]{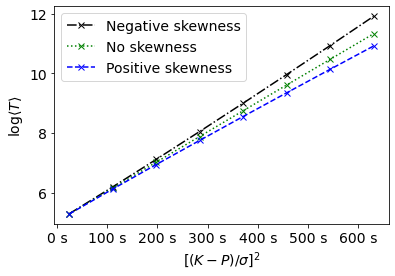}
    \caption{Impact of the skewness on the average desynchronization time $\langle T \rangle$, showing that a positive skewness in the noise entering in an area with positive power production $P$ decreases $\langle T \rangle$ and vice-versa. For further explanations see the text. }
    \label{fig: skewness}
\end{figure}

As we have derived in Sec.~\ref{secadditional},  a third cumulant (skewness) in the noise input can increase or decrease the time to desynchronization. What it actually does depends on where in the network it enters. As stated earlier, if it enters in the area with positive power input a positive (negative) skewness decreases (increases) the average time to desynchronization, while if it enters in the area with negative power input  a positive (negative) skewness increases (decreases) the average time to desynchronization.
A concrete example of this impact is shown in Fig.~\ref{fig: skewness}.
In the legend of Fig.~\ref{fig: skewness} 'positive skewness' refers to $\Gamma^{\xi}_3$ of Eq.~\ref{eq: third_cumulant}. The noise terms in areas $1$ (positive production) and $2$ (negative production),  $\xi_1$ and $\xi_2$ respectively, are constructed to have tuneable $\Gamma^{\xi}_3$. Similar to \cite{hindes_network_2019}, we construct compound Poisson processes according to Sec.~\ref{subsec31}. For the case of positive skewness we choose $\rho = 101$ and a jump size distribution
\begin{equation}
 \lambda_1(z) = \frac{100}{101}\delta \Big(z + \frac{1}{\sqrt{100 \cdot 1^2 + 1 \cdot 100^2}}\Big) + \frac{1}{101}\delta\Big(z - \frac{100}{\sqrt{100 \cdot 1^2 + 1 \cdot 100^2}}\Big)
\end{equation}
for $\xi_1$, and
\begin{equation}
\lambda_2(z) = \lambda_1(-z)
\end{equation}
for $\xi_2$. By Eqs.~\ref{eq: poiss_moments} and \ref{eq: z_from_xi_cumulants} this gives cumulants
\begin{eqnarray}
\Gamma_n^{\xi_1} &=& 100 \cdot \Big(\frac{-1}{\sqrt{100 \cdot 1^2 + 1 \cdot 100^2}}\Big)^n + 1 \cdot \Big( \frac{100}{\sqrt{100 \cdot 1^2 + 1 \cdot 100^2}}\Big)^n \nonumber \\
\Gamma_n^{\xi_2} &=& \Gamma_n^{\xi_1} \;\text{for even $n$} \nonumber \\
\Gamma_n^{\xi_2} &=& -\Gamma_n^{\xi_1} \;\text{for odd $n$}, \nonumber \\
\Gamma^{\xi}_n &=& \frac{1}{2^{n-1}}\big(\Gamma_n^{\xi_1} - \Gamma_n^{\xi_2}).
\end{eqnarray}
For the case of negative skewness we choose the same but with $\xi_1$ and $\xi_2$ swapped. The case of 'no skewness'  corresponds to the same choice but with $(1/\sqrt{2})\cdot (\xi_1 + \xi_2)$ entering (independently) at each node. The three cases have identical even cumulants $\Gamma^{\xi}_n$, and positive, negative, and zero odd cumulants respectively. Eq.~\ref{eq: third_cumulant} predicts the case with $\Gamma^{\xi}_3 > 0$ ($\Gamma^{\xi}_3 < 0$) to decrease (increase) $\langle T \rangle$ relative to the case with $\Gamma^{\xi}_3 = 0$, as is here confirmed by calculating $\langle T \rangle$ from the overload prediction of Eq.~\ref{eq: non-gaussian_action}, truncated after the first $8$ cumulants (and otherwise the same parameters as in Fig.~\ref{fig: escape_gauss}).

\section{Summary and Conclusions}\label{sec7}

We considered rare desynchronization events in a coarse-grained form of the Brazilian grid with prototypical data assignments for inertia, damping, power input and time correlations of the fluctuating production. As model we used the swing equations with stochastic power input that is supposed to reproduce histograms of measured power increment data with non-Gaussian features. We proposed a so-called Fourier-method to reconstruct the measured histograms that avoids inherent approximations in an alternative reconstruction by means of a compound Poisson process. For solar data both approaches lead to clear differences in the increment histograms and increment cumulants. Our approach provides a description of power fluctuations in terms of the cumulants of its increments (rather than its moments, as in the compound Poisson approach), whose convenient properties in summations allow for an understanding of non-Gaussian effects in heterogeneous power grids.\\
In view of large power grids, a dimensional reduction of the high-dimensional phase space is rather appreciated. Here we extended the so-called synchronized subgraph approximation \cite{hindes_network_2019} to the swing equations, arbitrary topologies and heterogeneous parameters. Although the partition into (subgraph) areas of the full grid is suggested by the values of the Fiedler vector at and close to the bifurcation point, the partition into disjoint synchronized areas after an overload remains applicable when a large fluctuation is necessary to induce the desynchronization far off the bifurcation point. \\
For really rare events, so rare that a sampling of the swing equations for the original or the reduced grid is unfeasible time-consuming, we applied the WKB-approach. The WKB-approach amounts to solving classical Hamilton's equations of motion. In accordance  with our different data implementation, we derive Hamilton's equations of motion directly in terms of (the derivative of) the cumulant generating function of the stochastic power increment input; this means for applying the WKB-approach, we have to calculate only the cumulants of the power increment input data. Due to the sudden drop of the involved auxiliary momenta when the desynchronization is due to an overload, the WKB-approach is only feasible when the phase space is further reduced. Via the subgraph approximation phase space gets reduced to twelve dimensions, and the solutions have been found by the iterative action minimization method. \\
A further reduction to two dimensions was based on the so-called overload approximation. This approximation applies when the optimal path between the stable and saddle fixed point, at which the system escapes in the WKB-approach, coincides with the time evolution of the phase difference between the two oscillators (areas) as a function of the power input. The trajectory of this phase difference follows the same equation that determines the fixed points, that is, which is obtained via dropping damping and inertia terms. Here an analytical solution can be found that gives the right order of magnitude of the average desynchronization time and agrees with results of the other approaches. \\
The action in the WKB approach is only determined up to a constant which is independent of the noise strength. For small enough noise strength or a sufficiently large distance to the bifurcation point, WKB is justified and all four approximations agree well with the results of the stochastic swing equations for the full grid.\\
For an increased contribution of renewables, an earlier expectation was to observe more frequent blackout events due to non-Gaussian tails of the power increments. Like \cite{hindes_network_2019} (but derived differently from \cite{hindes_network_2019}) we have shown that the skewness of the power increment distribution can lead to either more or less desynchronization events, depending on where in the grid the fluctuating input enters. Neither does lower inertia due to more renewables necessarily lead to more instabilities. As we have shown, for our realistic parameter values, the order of magnitude of the average time to desynchronization is independent of inertia and damping. Inertia and damping should not be considered in isolation; what matters is their relation to the time correlations of the fluctuations. Tuning the correlation time $1/\alpha$ amounts to tuning the inertia and damping.

In general, the role of time correlations in the fluctuating power production is not negligible. Our results have been obtained for $1/\alpha=100-300 \;\text{s}$, corresponding to measured correlations in \cite{tyloo_finite-time_2022}. It is $\alpha$ that determines the degree to which the results are impacted by non-Gaussianity of the fluctuating input. Along with the strength of the noise $\sigma$ it enters the action in the WKB-approach. Setting the correlation time $1/\alpha=0 \;\text{s}$ as for white noise is definitely not justified for typical wind or solar data. Further work should therefore extend this investigation of the interplay between non-Gaussianity and finite time-correlations also to account for more realistic power spectra \cite{katrin}.

Due to the exponential sensitivity of the average time to desynchronization, small differences in parameters may get strongly amplified in the very prediction of how rare blackouts are. Therefore care is needed with respect to results, for which parameters or data are homogenized for simplicity.

\section*{Acknowledgment} We thank Jason Hindes (U.S. Naval Research Laboratory) and Bhumika Thakur (Jacobs University Bremen) for useful discussions. Financial support of the Bundesministerium f\"ur Bildung und Forschung (BMBF) (grant number 03EK3055D) is gratefully acknowledged. \\

\bibliography{library}

\appendix
\section{Parameters for the reduced Brazilian grid}
The data (\cite{birchfield_grid_2017}) contains the topology of the grid, the reactance for each line (equal to $1/K_{ij}$ when resistance is neglected, as is the case for the swing equations) and the generators and loads connected to each node, along with the production/consumption at these nodes. The data contains a transformer, with a node for each end. As no (relevant) data for the transformer is given, and we cannot model the transformer within the swing equations (without voltage dynamics), we treat this as a single node. For synchronous generators and synchronous loads the constants of inertia $H_i$ are given, from which $M_i = \frac{2H_i}{\omega} |P_i|$ can be determined (where $|P_i|$ is the production/consumption at the generator/load). For the Brazilian grid, the angular frequency $\omega = 2 \pi \cdot 60 \;\text{Hz}$. For these generators and loads, we assume (as argued in \cite{nishikawa_comparative_2015}) that the damping coefficient is dominated by a droop of $2 \%$ (see \cite{machowski_power_2008} for a detailed description of droop). The damping of inductive generators and loads is due to a small self-regulating effect of $\approx 1 \%/\text{Hz}$ \cite{UCTE}; for these devices the inertia is set to zero. If there are multiple generators/loads connected to a node, we sum their inertia, damping, power input. One of the nodes does not have any inertia; at this node we set the inertia to a small value ($0.01\,\text{s}^2$) to avoid numerical instability.

This gives the following sets of parameters, here shown at $9.1 \%$ from overload, in the per unit system (i.e. in units of a reference power that itself is here not further specified, see \cite{nishikawa_comparative_2015, machowski_power_2008}):
\begin{alignat}{1}
    \bm{P} &= \begin{pmatrix} -1.38304 & 1.17679 & 1.56501 & 12.38545 & -1.05426 & -12.68997 \end{pmatrix} \,,\\
    \bm{M} &= \begin{pmatrix} 0.06201 & 0.04950 & 0.05716 & 0.29119 &  0.01 & 0.14902 \end{pmatrix}\,\text{s}^2 \,,\\
    \bm{\gamma} &= \begin{pmatrix}0.35049373 & 0.27670489 & 0.31923646 & 1.43640985 & 0.11831416 & 0.75322515
    \end{pmatrix}\,\text{s}\,,\\
    \bm{K} &= \begin{pmatrix} 0. & 0.& 26.316 & 0. & 4.082 & 0. \\
                         0. & 0.& 13.158 & 0. & 4.444 & 0.\\
                        26.316 & 13.158&  0.&  0.& 0. & 0. \\
                        0.&  0. & 0. & 0. & 13.624 &0. \\
                        4.082 & 4.444 & 0. & 13.624 & 0. & 17.544 \\
                        0. & 0.& 0.& 0. &17.544& 0. \end{pmatrix} \,.
\end{alignat}
\section{Derivation of the differential Chapman-Kolmogorov equation}\label{subsubsec43}
Under mild assumptions, further specified in \cite{gardiner_stochastic_2009}, for any continuous time Markov process, here for the swing equations with fluctuations according to Eq.~\ref{eq: p_dynamics}, the time evolution of  $\pr(\bm{\phi}, \bm{v}, \bm{p}, t)$
is given by an integro-differential equation called the differential Chapman-Kolmogorov equation \cite{gardiner_stochastic_2009}. For $\pr(\bm{\phi}, \bm{v}, \bm{p}, t)$  this equation can be derived from Eqs.~\ref{eq: swing_dynamics}-\ref{eq: p_dynamics} by first considering the discrete evolution of $\pr(\bm{\phi}, \bm{v}, \bm{p}, t)$:
\begin{alignat}{1}
    \frac{\pr(\bm{\phi}, \bm{v}, \bm{p}, t + \Delta t) - \pr(\bm{\phi}, \bm{v}, \bm{p}, t)}{\Delta t} &=\frac{1}{\Delta t}\int \dd \bm{\phi}' \dd \bm{v}' \dd \bm{p}' \label{eq: discrete_m_eq}\\
    \times \Big[ \pr(\bm{\phi}, \bm{v}, \bm{p}, t + \Delta t|\bm{\phi}', \bm{v}', \bm{p}', t)\pr(\bm{\phi}', \bm{v}', \bm{p}', t) &- \pr(\bm{\phi}', \bm{v}', \bm{p}', t + \Delta t|\bm{\phi}, \bm{v}, \bm{p}, t) \pr(\bm{\phi}, \bm{v}, \bm{p}, t) \Big] \,.\nonumber
    \end{alignat}
For a given realization of the noise $\xi$, the transition probabilities $\pr(\bm{\phi}, \bm{v}, \bm{p}, t + \Delta t|\bm{\phi}', \bm{v}', \bm{p}', t)$ and $\pr(\bm{\phi}', \bm{v}', \bm{p}', t + \Delta t|\bm{\phi}, \bm{v}, \bm{p}, t)$ are simply delta functions enforcing the swing equations (Eqs.~\ref{eq: swing_dynamics}-\ref{eq: p_dynamics}). For small $\Delta t$, we get,
\begin{alignat}{1}
    &\frac{1}{\Delta t}\pr(\bm{\phi}, \bm{v}, \bm{p}, t + \Delta t|\bm{\phi}', \bm{v}', \bm{p}', t) \nonumber\\
    &= \frac{1}{\Delta t} \prod_i \Big[\int \dd z_i \pr(Z_{\Delta t, i} = z_i) \times \delta \Big( (\phi_i - \phi_i') - \Delta t \cdot v_i \Big) \nonumber\\
    &\times \delta\Big((v_i - v_i') - \frac{\Delta t}{M_i} \cdot \big[-\gamma_i v_i' + \overline{P}_i + p_i' + \sum_j K_{ij} \sin(\phi_j - \phi_i) \big] \Big) \nonumber\\
    &\times \delta \Big((p_i- p'_i) - [- \alpha_i p_i' \Delta t  + \sigma_i z_i] \Big) \Big] \,,
\end{alignat}
and equivalent expressions for $\pr(\bm{\phi}', \bm{v}', \bm{p}', t + \Delta t|\bm{\phi}, \bm{v}, \bm{p}, t)$. Expanding the delta functions to first order in $\Delta t$ (with the derivative of the delta function defined such that $\int \dd x f(x) \frac{\text{d}\delta(x)}{\text{d}x} = - \frac{\dd f(x)}{\dd x}\biggr \rvert_{0}$), the limit $\Delta t \rightarrow 0$ leads from Eq.~\ref{eq: discrete_m_eq} to:
\begin{alignat}{1}
    \frac{\partial \pr(\bm{\phi}, \bm{v}, \bm{p}, t)}{\partial t} = \sum_i \Big[ &- \frac{\partial}{\partial \phi_i}[v_i \cdot \pr(\bm{\phi}, \bm{v}, \bm{p}, t)] \nonumber \\
    &- \frac{\partial}{\partial v_i}\big[\frac{1}{M_i}\big(- \gamma_i v_i + \overline{P}_i + p_i + \sum_j K_{ij} \sin(\phi_j - \phi_i)\big) \cdot \pr(\bm{\phi}, \bm{v}, \bm{p}, t) \big]] \nonumber\\
    &- \frac{\partial}{\partial p_i} \big[-\alpha_i p_i \cdot \pr(\bm{\phi}, \bm{v}, \bm{p}, t)\big] \nonumber\\
    &+ \lim_{\Delta t \rightarrow 0} \int \dd z_i \frac{\pr(Z_{\Delta t, i} = z_i)}{\Delta t} \cdot \big(\pr(\bm{\phi}, \bm{v}, \bm{p} - \bm{\mathbbm{1}_i} \sigma_i z_i, t) - \pr(\bm{\phi}, \bm{v}, \bm{p}, t) \big) \Big] \,,
\end{alignat}
where $\bm{p} - \bm{\mathbbm{1}}_i \sigma_i  z_i \equiv (p_1, \dots, p_{i-1}, p_i -  \sigma_i z_i, p_{i + 1} \dots)$. The last term can be extended according to:
\begin{alignat}{1}
    \lim_{\Delta t \rightarrow 0}\int \dd z_i \Big(\frac{\pr(Z_{\Delta t, i} = z_i)}{\Delta t} + (1 - \frac{1}{\Delta t }) \delta(z_i) \Big) \cdot \big(\pr(\bm{\phi}, \bm{v}, \bm{p} - \bm{\mathbbm{1}_i}\sigma_i z_i, t) - \pr(\bm{\phi}, \bm{v}, \bm{p}, t) \big)\, .
\end{alignat}
After defining the transition rates
\begin{alignat}{1}
    w_i(z_i) &\equiv \lim_{\Delta t  \rightarrow 0} \Big(\frac{\pr(Z_{\Delta t, i} = z_i)}{\Delta t} + \big(1 - \frac{1}{\Delta t}) \delta(z_i) \Big) \,\label{eq:wiz}
\end{alignat}
this then gives the evolution of the probability distribution $\pr(\bm{\phi}, \bm{v}, \bm{p}, t)$ according to Eq.~\ref{eq: m_eq}.
The additional terms $(1- \frac{1}{\Delta t})\delta(z_i)$ ensure that $\int \mathrm{d}z_i \, w_i(z_i) = 1$, that $w_i(z_i)$ converges to a distribution for $\Delta t \rightarrow 0$ and that it has a finite moment-generating function $\langle \exp(J z_i) \rangle_{w_i(z_i)}$.
The moment generating function of $w_i(z_i)$ can be found by using that from Eq.~\ref{eq: z_from_xi_cumulants} the cumulant generating function of $Z_{\Delta t, i}$ is related to that of $\xi_i$ by $\log \langle \exp(J z_i) \rangle_{\pr(Z_{\Delta t, i} = z_i)} = \Delta t \, C^{\xi_i}(J)$:
\begin{alignat}{1}
    \langle \exp(J z_i) \rangle_{w_i(z_i)} &= \lim_{\Delta t  \rightarrow 0} \Big( \big[\langle \exp(J z_i) \rangle_{\pr(Z_{\Delta t, i} = z_i)}\big]/\Delta t + \big(1 - \frac{1}{\Delta t}) \Big) \,. \nonumber\\
    &= \lim_{\Delta t  \rightarrow 0} \Big(\exp \big[\Delta t C^{\xi_i}(J)\big] /\Delta t + \big(1 - \frac{1}{\Delta t}) \Big) \,. \\
    &= 1 +  C^{\xi_i}(J) \,, \label{eq: w_mgf}
\end{alignat}
and hence the moment generating function of $w_i(z_i)$ equals (up to a constant) the cumulant generating function $C^{\xi_i}(J) = \Gamma_1^{\xi_i} \frac{J}{1!} + \Gamma_2^{\xi_i} \frac{J^2}{2!} + \dots $ of $\xi_i$. For $n \geq 2$ the moments of $w_i(z_i)$ (and hence the cumulants of $\xi_i$) correspond to the Kramers-Moyals coefficients (see Sec.~\ref{subsec31}), here independent of $p_i$:
\begin{alignat}{1}
 M^{(n)}_i \equiv \lim_{\Delta t \rightarrow 0}\frac{1}{\Delta t}\langle \big(p_i(t + \Delta t) - p_i(t) \big)^n\rangle &= \lim_{\Delta t \rightarrow 0}\frac{1}{\Delta t}\langle \big(\sigma_i Z_{\Delta t, i}\big)^n\rangle \nonumber \\
 &= \sigma_i^n \int \text{d}z_i \, z_i^n \, w(z_i) = \sigma_i^n \Gamma_n^{\xi_i} \,,
\end{alignat}
where in the second equality sign we used Eq.~\ref{eq: Z_from_data}. Closely related to the moment generating function of $w_i(z_i)$ is its Fourier transform, found from Eqs.~\ref{eq:24} and \ref{eq:wiz} to be equal to:
\begin{alignat}{1}
    \mathcal{F}\big(w_i \big)[\omega] &= 1 + \frac{1}{\tau} \Big[\Log \mathcal{F}\big(\pr(Z_{\tau,i})\big)[\omega] + n(\omega) \cdot 2 \pi i \Big] \,. \label{eq: w_from_fourier}
\end{alignat}
For a data implementation according to the Fourier-method we have $\pr(\sigma_i Z_{\tau,i}) = \pr(\text{increment data})$ (Eq.~\ref{eq: data_identification}), which can in principle be used to numerically calculated $w_i(z_i)$ from the increment data.
On the other hand, when $\xi_i$ follows the compound Poisson process construction of Sec.~\ref{subsec31}, then:
\begin{alignat}{1}
\pr(Z_{\Delta t, i} = z_i) &= \pr(Z_{\Delta t, i} = z_i | \text{no jumps})\cdot \pr( \text{no jumps}) \nonumber \\
&+ \pr(Z_{\Delta t, i} = z_i | \text{one jump})\cdot \pr( \text{one jump}) + \dots  \nonumber \\
&= \delta(z_i) \cdot \frac{(\rho \Delta t)^0 \exp(- \rho \Delta t)}{0!} + \lambda(z_i) \cdot \frac{(\rho \Delta t)^1 \exp(- \rho \Delta t)}{1!} + \dots  \nonumber \\
&= \delta(z_i) \cdot (1 - \rho \Delta t ) + \rho \lambda(z_i) \cdot \Delta t + O\big([\Delta t]^2\big)\,,
\end{alignat}
which gives $w_i(z_i) = \rho \lambda_i(z_i) + (1 - \rho) \delta(z_i)$.
\end{document}